\documentclass{aastex6}

\usepackage{natbib}
\usepackage{xcolor}

\newcommand{\eps}[1]{\mbox{log~$\epsilon$(#1)}} 
\newcommand\species[2]{#1 {\sc #2}}

\def\ie{\mbox{i.e.}}
\def\eg{\mbox{e.g.}}

\def\teff{\mbox{T$_{\rm eff}$}}
\def\logg{\mbox{log~{\it g}}}
\def\vmicro{\mbox{$\xi_{\rm t}$}}
\def\kmsec{\mbox{km~s$^{\rm -1}$}}

\shorttitle{Comparison of MR and MP RRab stars}
\shortauthors{Chadid et al.}

\begin{document}

\title{Spectroscopic Comparison of Metal--Rich RRab Stars 
of the Galactic Field with Their Metal--Poor Counterparts}

\author{
        Merieme Chadid\altaffilmark{1},
        Christopher Sneden\altaffilmark{2}, and 
        George W.Preston\altaffilmark{3},
}

\altaffiltext{1}{Universit\'e Nice Sophia--Antipolis, Observatoire de la C\^ote d’Azur, 
                 UMR 7293, Parc Valrose, F-06108, Nice Cedex 02, France; 
                 chadid@unice.fr}
\altaffiltext{2}{Department of Astronomy and McDonald Observatory, The University of 
                 Texas, Austin, TX 78712, USA; chris@verdi.as.utexas.edu}
\altaffiltext{3}{Carnegie Observatories, 813 Santa Barbara Street, Pasadena, CA 91101, USA;
                 gwp@obs.carnegiescience.edu}

\begin{abstract}

We investigate atmospheric properties of 35 stable RRab 
stars that possess the full ranges of period, light amplitude, and metal 
abundance found in Galactic RR Lyrae stars. 
Our results are derived from several thousand echelle spectra obtained 
over several years with the du Pont telescope of Las Campanas Observatory.
Radial velocities of metal lines and the H$\alpha$ line were used to
construct curves of radial velocity $versus$ pulsation phase.  
From these we estimated radial velocity amplitudes for 
metal lines (formed near the photosphere) and H$\alpha$ Doppler cores 
(formed at small optical depths).  
We also measured H$\alpha$ emission fluxes when they appear during primary 
light rises.  
Spectra shifted to rest wavelengths, binned into small phase intervals,
and coadded were used to perform model atmospheric and abundance analyses.
The derived metallicities and those of some previous spectroscopic surveys 
were combined to produce a new calibration of the Layden abundance scale.
We then divided our RRab sample into metal-rich (disk) and metal-poor 
(halo) groups at [Fe/H] = $-$1.0,
The atmospheres of RRab families, so defined, differ with respect to 
(a) peak strength of H$\alpha$ emission flux, (b) H$\alpha$ radial velocity 
amplitude, (c) dynamical gravity, (d) stellar radius variation, (e) secondary 
acceleration during the photometric “bump” that precedes minimum light, 
and (g) duration of H$\alpha$ line-doubling.  
We also detected H$\alpha$ line-doubling during the ``bump'' in the 
metal-poor family, but not in the metal-rich one.  
Though all RRab probably are core helium-burning horizontal branch stars, 
the metal-rich group appear to be a species sui generis.

\end{abstract}


\section{INTRODUCTION}

Because of their moderately high luminosities, $M_V$ $\sim$ 0.5 \citep{Arp59}
and plentiful numbers in globular clusters (hereafter GC) and in the 
Galactic field, much effort has been expended to calibrate the RR Lyrae 
stars as standard candles, both observationally, e.g., \cite{Kollmeier13} and 
theoretically, e.g. \cite{marconi15} in order to explore the spatial and 
kinematic structures of the Galactic halo and its attendant satellite systems. 
Bailey's (1902)\nocite{bailey02} subtypes RRa and RRb exhibit periodic 
pulsations (0.4 $<$ $P$ $<$ 0.9~d) in the fundamental mode with strongly 
skewed light variations of large amplitude ($\sim$1 visual magnitude) that 
are readily detected by conventional photometric methods. 

The RRab stars of the Galactic field possess a large range in metallicity 
\citep{preston59,layden94}, including a metal-rich component with disk spatial 
distribution and kinematic 
properties \citep{preston59,layden95a,layden95b}
that has no counterpart among the Galactic GCs.  
Metallicities as high as solar ([Fe/H]~=~0)\footnote{
We adopt the standard spectroscopic notation \citep{wallerstein59} that for
elements A and B,
[A/B] $\equiv$ log$_{\rm 10}$(N$_{\rm A}$/N$_{\rm B}$)$_{\star}$ $-$
log$_{\rm 10}$(N$_{\rm A}$/N$_{\rm B}$)$_{\odot}$.
We use the definition
\eps{A} $\equiv$ log$_{\rm 10}$(N$_{\rm A}$/N$_{\rm H}$) + 12.0, and
equate metallicity with the stellar [Fe/H] value.}
occur and the fundamental periods are systematically smaller than those 
of more metal-poor variables. 
In particular the Galactic field contains metal-rich RRab stars with large 
light amplitudes, $\Delta V$~$>$~1, and periods less 
than 0.43~d that have never been found in any GC. 
RRab stars with these characteristics are more concentrated toward the Galactic 
plane than RRab stars of longer period.
This metal-rich short-period extension in the Galactic field is historically 
the first metallicity effect encountered in the RRab family.  
\cite{kukb49} first called attention to this anomaly more than 65 years ago.  
How the metal-rich RRab stars should be apportioned between the thick 
disk \citep{gilmore83} and an old thin disk (\citealt{preston91}, 
\citealt{layden94}, \citealt{morrison90}) has yet to be determined.  

In Figure~\ref{f1}, adapted from Figure~1 of \cite{layden95b} we present 
a plot of pulsation period $versus$ [Fe/H] for the RRab stars 
tabulated by (\citealt{layden94}, hereafter called Lay94). 
Periods were taken from the General Catalog of Variable Stars (GCVS;         
\citealt{sam10}\footnote{http://www.sai.msu.su/gcvs/gcvs/},   
except for metal--rich CI~And for which we used the period derived by        
\citep{schmidt91}. 
We adopt a division between metal-poor (blue circles) and metal-rich 
(red circles) stars at [Fe/H]~= $-$1.0 in view of the abrupt changes in 
kinematic properties at this metallicty
assembled in Layden's (1995a)\nocite{layden95a} Figures 1, 3, and 4.  
We are mindful of the $bona fide$ overlap of the abundance distributions 
of the Galactic halo and disk (\citealt{morrison90}, \citealt{ruchti11}).  
Regrettably, the small data sample of the present investigation is not 
well-suited to investigate such overlap in the RRab populations.

The trend in Figure~\ref{f1} toward shorter periods 
with increasing [Fe/H] is accompanied by a modest decrease in luminosity 
($\sim$1 magnitude) that has been the subject of numerous investigations 
summarized by \cite{mcnamara99}. 
Conclusions about the nature of the metal rich component are insensitive 
to the precise location of the disk-halo abundance division point at the 
0.1~dex level. 
\cite{zinn85} placed the boundary between disk and halo GCs at [Fe/H]~= $-$0.8;
a choice that would not materially affect the present study.

The great majority of Zinn's (1985)\nocite{zinn85} disk GC have red 
horizontal branches that contain few, if any, RRab variables. 
The one RRab star in the metal-rich globular cluster 47~Tuc has a long 
period of 0.737 d \citep{carney93}.
Contrary to all expectations \cite{rich97} reported that two metal-rich GCs, 
NGC~6388 and NGC~6441, possess extended blue HBs, and subsequent searches 
for variable stars (\citealt{pritzl02}, \citealt{corwin06}) revealed the 
presence of numerous RRab stars with unexpectedly long periods. 
The period distributions of the metal rich field and cluster variables 
barely overlap, as is evident from a glance at Figure~\ref{f1}.
Clearly, the metal-rich RRab stars of the Galactic field 
and those in Zinn’s disk GCs are not drawn from the same parent population,
an anomaly recognized already by \cite{layden95b}.
The reason for the different period distributions of the field and GC 
metal-rich RRab populations is unknown at present.
We suppose that a complete explanation will address the existence of multiple 
stellar populations in these two GCs \citep{bellini13} and will appeal 
to model calculations of the sort made by \cite{bono97a,bono97b} and 
\cite{marconi15}.  
An early, semi-empirical explanation of metal-rich RRab stars by 
\cite{taam76} required a special HB evolutionary scenario, preceded by very 
large ($\sim$0.5$M_\odot$), unverified mass loss during RGB evolution.  
Though never discredited, their proposal is not widely accepted, and the 
search for antecedents of metal-rich RRab stars continues.

Although a multitude of fascinating photometric characteristics of the 
RR Lyrae stars have been investigated intensively in recent times, with 
particular emphasis on the mysterious \cite{bla07} phenomenon, the 
relatively few high-resolution spectroscopic surveys have been directed 
primarily toward derivation of elemental abundances and toward 
luminosities via the Baade-Wesselink method. 
We recognize the interest that attends the Blazhko mystery.  
However, from the perspective of this paper the Blazhko phenomenon 
only complicates an already overburdened topic.  
We limit our discussion to stable RRab stars as defined by \cite{chadid13}, 
deferring Blazhko issues to a subsequent investigation. 

This paper is devoted to a detailed high-resolution spectroscopic comparison 
of samples of nearby metal-rich and metal-poor RRab stars of the Galactic 
field that have been observed at Las Campanas Observatory during the past 
decade (see \citealt{preston11} \S5.1).  
We derive stellar atmospheric parameters effective 
temperature, microturbulent velocity, and gravity as functions of 
pulsation phase, [Fe/H] metallicity, and period.  
Radial velocities, derived separately from metal lines and from the H$\alpha$ 
line of hydrogen, are used to explore changes in the atmospheric 
structures of these stars that occur during their pulsation cycles.   
We place extant metallicity results from several surveys on a common system, 
and derive relative abundance ratios of several elements.
Then we compare the kinematic behaviors of metal-rich and metal-poor RRab 
envelopes.  
We find significant differences between the envelope kinematics of RRab 
abundance families, 
\textit{families that are defined solely by their Galactic kinematics} 
(Layden's 1995a\nocite{layden95a} Figures 1, 3, and 4).  
This surprises us.
Insofar as we are aware these differences play no role in either the 
construction or interpretation of horizontal branch models.  
They are phenomena sui generis.

\section{OBSERVATIONS AND REDUCTIONS}

The survey of RRab stars upon which this paper rests was conducted with 
the echelle spectrograph of the du Pont telescope.
The spectrograph was configured in the manner used for our previous RRab 
and RRc metallicity studies {\citep{for11a,for11b,govea14}}.  
Specifically, the instrument was used with a 1.5 x 4.0\,arcsec aperture
which yielded a resolving power 
$R$~=~$\lambda/\Delta\lambda$ $\sim$\,27000 at 5000\,\AA\
and wavelength coverage $\lambda\lambda$ 3400--9000.

Observations were gathered in the eight-year interval 2006--2013.  
In 2006--2009 ten RRab stars with periods near 0.57\,d were observed to 
provide a basis for comparison of s-process rich TY\,Gru {\citep{preston06}} 
with RRab stars of normal halo composition.  
Following discovery in 2009 of emission and absorption lines of \species{He}{i} 
and \species{He}{ii} in AS\,Vir the survey was extended to 24 additional 
stars that sample the entire ranges of period, light amplitude, and 
metallicity encountered in the RRab family {\citep{preston09,preston11}}.  
These additional stars, listed in Table~\ref{tab-stars}, were observed 
intensively during twenty 5- to 7-night runs during the years 2009–-2013.  

The photometric periods of our program stars range from 0.39 to 0.71\,days, 
during which they vary in brightness with light amplitudes on 
$0.63 < V_{amp} < 1.28$ and highly correlated metal radial velocity 
amplitudes, $50 < \Delta\,RV < 70$~\kmsec (Table\,\ref{tab-stars}).  
We limited our exposure times to a maximum of 600\,s. 
More typically the exposures were 400\,s, with some as short as 200\,s.
The maximum integration times were thus less than 2\% of the shortest 
pulsation period among our program stars. 
This procedure rendered the $RV$ variations that occur during integrations 
much smaller than other spectral line broadening effects in individual 
observations.
However, the integration-time restriction resulted in low signal-to-noise 
individual spectra.

Note that particular emphasis was placed on primary light rises due to initial 
interest in the shock phenomena that produce hydrogen and helium emission.  
We have come to regret this tactic, because it led to inadequate phase 
coverage of phenomena during declining light that we only became aware 
of late in the survey.

We extracted spectra from the original CCD image files using procedures 
discussed previously \citep{preston00,for11a}; for the most part the
description of these reduction steps need not be repeated here.  
However, we have adopted one shortcut in the removal of scattered light 
from our spectra that deserves discussion inasmuch as it bears on the 
accuracy of abundances derived in this paper and their use in recalibration 
of the RR~Lyrae abundance scale.

\subsection{Removal of Scattered Light from du Pont echelle spectra\label{scat}}

\cite{for11b} describe a scattered light problem peculiar to the du Pont 
echelle spectrograph, and they outline a procedure used to remove scattered 
light from their observations.  
This special technique is required because the projected length of the 
entrance aperture of the du Pont echelle, constrained by prismatic order 
separation, is too small longward of 5000\,\AA\ to permit 
subtraction of scattered light by use of pixels adjacent to the projected 
stellar image.  
In this paper we use a greatly simplified procedure, namely, subtraction of 
a constant fraction of the stellar flux at each pixel along each echelle order.
This model attributes all scattered light to small-angle scattering into 
regions immediately adjacent to each spectral order.  
It ignores large-angle scattering that could, for example, flood high 
(blue-violet) orders with red light from distant low (red) spectral orders 
in the echelle format. 
Large-angle scattering into blue-violet orders can be important in the case 
of cool, red stars, but RR Lyrae are not very red, so our simplified correction 
works well.  
We demonstrate this with measurements of equivalent widths ($EW$) in the 
spectrum of one of our radial velocity standard stars, the metal--poor 
subgiant HD\,140283. 
This star is somewhat redder ($B-V$~=~0.50) than the RRab stars, for which 
colors lie in the range 0.1~$< B-V <$~0.5 during their pulsation cycles.

We added 38 du Pont echelle spectra of HD\,140283, flattened them, 
subtracted 0.1 from the normalized continua in all orders, and renormalized.
The $S/N$ in this spectrum increases from 250 at 4000\,\AA\ to 
500 at 5500\,\AA.
For this test we used the IRAF/splot\footnote{
IRAF is distributed by the National Optical Astronomy Observatory, which    
is operated by the Association of Universities for Research in Astronomy    
(AURA) under cooperative agreement with the National Science Foundation.} 
 package to measure $EW$s of the 
\species{Fe}{i} and \species{Fe}{ii} lines that were measured in a Suburu 
spectrum of HD\,140283 by \cite{Gallagher10}.
We correlate these two $EW$ sets in Figure\,\ref{f2}, plotting the Suburu
values (labeled EWS) $versus$ those from du Pont (EWD) in panel (a), and 
line-by-line differences EWS~$-$~EWD as functions of line strength EWS
in panel (b) and wavelength in panel (c).
The agreement is good: $\langle$EWS~$-$~EWD$\rangle$ = $-$0.48~m\AA,
with standard deviation $\sigma$ = 1.01\,m\AA\ from 83 lines in common.

The $EW$ differences in panel (c) of the figure allow us
to investigate whether large angle scattering from the brighter low (red) 
orders in the du Pont spectrograph produces wavelength-dependent residuals.
The lack of significant variation with wavelength leads us to conclude
that our scattered light corrections are acceptable, \ie, they play an 
insignificant role in the error budget of our abundance analyses.

\section{MODEL ATMOSPHERES AND CHEMICAL COMPOSITIONS\label{atmos}}

We derived model atmosphere parameters effective temperature \teff, 
surface gravity \logg, metallicity [Fe/H] and microturbulent velocity \vmicro\
for each program star at many pulsational phases, following the procedures 
discussed by {\cite{{for11b}} and {\cite{{govea14}}. 
The integration time restrictions outlined in \S2 yielded
low signal-to-noise spectra, frequently $S/N < 20$.
To increase S/N we created co--added spectra in the following manner.
For each star we sorted all available spectra in order of increasing 
pulsation phase by use of the elements in Table\,\ref{tab-stars}.
We divided these sorts into phase bins.  
The bounds of these phase intervals chosen for co-addition were dictated 
in most instances by vagaries of the phase distributions.  
The phase intervals of the bins were generally smaller than 0.05\,P and 
never larger than 0.10\,P.  
The number of spectra per bin ranged from 3 to 10.  
We excluded observations made on $0.6 < \phi < 0.9$ because of large 
macroturbulent velocities that degrade the spectra in this phase interval 
\citep{Preston14}.
Each spectrum in each bin was shifted to rest wavelength with IRAF/dopcor
by use of radial velocities measured in IRAF/fxcor.  
Finally, average spectra were created from these shifted spectra by
use of IRAF/scombine.

\subsection{Model Atmosphere and Abundance Calculations\label{analysis}}

We used the co-added spectra to first derive their model atmospheric 
parameters.
We also decided to re-derive models and abundances for the 10 stars 
originally discussed by \cite{for11b}, in order to ensure that their
parameters were on the same system that we employed for the rest of our sample.
We used spectra of one co-added phase bin at $\phi = 0.5$ for their stars. 
Here we highlight the main analytical points and note any substantive 
differences between the present and previous studies.

We adopted the line list of \cite{for11b}, updating the transition 
probabilities with new laboratory values for \species{Ti}{i} 
\citep{Lawler13}, \species{Ti}{ii} \citep{Wood13}, and \species{Fe}{i} 
\citep{Den14}, combining these with values from \cite{brian91} for lines 
with excitation energies $\chi$~$<$ 2.3~eV. 
For these lines we measured $EW$s using an Interactive Data Language 
(IDL)\footnote{
IDL is proprietary software system distributed by Exelis Visual Information
Solutions, Inc., and is available at
http://www.exelisvis.com/ProductsServices/IDL.aspx.} 
code developed by \cite{Roederer10} and \cite{Brugamyer11}. 
These line lists and interpolated ATLAS (\cite{kurucz11} and references 
therein) model atmospheres were used as inputs to the current version of 
the $LTE$ line analysis code $MOOG$ {\citep{Sneden73}}\footnote{
Available at http://www.as.utexas.edu/$\sim$chris/moog.html.}.
In the analyses we eliminated those lines that are too weak for reliable 
measurement.

The results of these calculations were line-by-line abundances for all species.
The parameters of the models were altered and the line analyses repeated to 
derive: (a) \teff, so that there be no average trend of abundances with 
excitation energy derived from \species{Fe}{i} lines; (b) \vmicro, so that 
there be no average trend of abundances from \species{Fe}{i} with $EW$; 
(c) \logg, so that mean abundances derived from \species{Fe}{i} and 
\species{Fe}{ii} agree well, and secondarily the mean abundance derived 
from \species{Ti}{i} not be in severe disagreement with that from 
\species{Ti}{ii}; and (d) [M/H], so that the assumed model metallicity was 
close to the derived [Fe/H] value. 
This procedure took typically three or fewer iterations. 
The rapidity of convergence was a result of two RRab properties. 
First, within the confines of traditional analyses, the 
RRab stars occupy a relatively small region of model parameter space, 
typically 6000 $< \teff\ <$ 7000\,K, 1.7 $< \logg\ <$ 2.7, 
and 2.5 $< \vmicro\ <$ 3.5\,\kmsec. 
Second, these stars are warm enough that the free electrons needed for the 
dominant H$^-$ continuous opacity are largely donated from H itself instead of 
the easily ionized metals, and so the widely-ranging metallicities 
($-$2.5 $<$ [M/H] $<$ 0.0) do not severely affect the model atmosphere 
structures.

We also derived several abundance ratios [X/Fe] for each program star at 
each pulsational phase. 
However, abundance ratios are of secondary interest in this paper, so given 
the $S/N$ limitations of our spectra we have chosen to present in 
Table\,\ref{tab-abunds} mean abundance ratios [X/Fe] for each star only for 
species that were available for analysis at most phases of most stars: 
\species{Ca}{i}, \species{Sc}{ii}, \species{Ti}{i} \& {\sc ii},
\species{Cr}{i} \& {\sc ii}, \species{Mn}{i}, and \species{Ba}{ii}, along with 
the mean [Fe/H] metallicities. 
All abundance ratios are referenced to the solar photospheric values 
recommended in the {\cite{asplund09}} review.  
Those values are in many cases based on analytical assumptions (\eg, non-LTE,
multi-stream modeling) that are beyond the scope of our standard abundance
analysis techniques.
Recently, a series of papers \citep{gre15,sco15a,sco15b} has updated the
solar abundances, computing them in a variety of ways.  
The solar abundance differences they derive are typically not large, usually
less than 0.05~dex for the species of interest here.
These are small uncertainties compared to the other error sources in our work.

The abundance ratios were computed for the same ionization-state abundances in 
numerator and denominator, \eg, [Ca/Fe] from \species{Ca}{i} and 
\species{Fe}{i}, and [Sc/Fe] from \species{Sc}{ii} and \species{Fe}{ii}.
These ratios have been computed from those derived at each phase for a star. 
At the bottom of this table we also list the mean line-to-line abundance 
scatters for each species $\langle\sigma\rangle$ and the mean number of 
lines employed in the calculations $\langle n \rangle$ in individual 
star/phase abundance derivations. 
These numbers can vary from star to star and phase to phase, but suggest 
that the $\sigma$ values never are less than $\simeq$0.1. 
Additionally, some species are represented by three or fewer transitions. 
Such abundance results should not be over-interpreted.

Final model atmosphere parameters for the program stars in all co-added 
phase bins are listed in Table\,\ref{tab-models}. 
In considering these values the reader should keep in mind that they 
result from application of static, plane-parallel model atmospheres and 
LTE line analysis techniques to stellar atmospheres that are highly dynamic. 
In less than a day the RRab stars cycle through correlated variations of 
1000\,K in \teff\ and 1\,dex in \logg and \vmicro\ (e.g., \citealt{for11b}).  
The derived spectroscopic gravities have static and dynamic components.
Moreover, the derived microturbulent velocities reflect both the 
standard small-scale turbulent velocities that are applied in analyses of 
non-variable stellar atmospheres and radial velocity gradients within the 
line forming regions of RR Lyrae pulsators. 

The derived absolute quantities should be viewed with caution.  
Relative values from star to star among our RRab sample should be much 
more reliable. 
All spectral features studied in this paper are due to ``photospheric'' 
absorptions by $\alpha$- and Fe-group elements in a restricted strength 
($EW$) range. 
Thus their lines are formed in similar atmospheric layers, and their
abundance ratios should be less sensitive to modeling uncertainties than 
are the individual abundances.

\subsection{Correlations of Model Parameters with Pulsation Phases and
            Metallicities\label{modelcorr}}

The derived temperatures of our stars vary with phase in a manner broadly 
consistent with the results of \cite{for11a}.  
Immediately after maximum light the temperatures decline, rapidly at first 
and then more slowly from phase $\sim$\,0.3 to phase 0.6, the upper phase 
limit of our binned spectra.  
The \teff\ values for individual stars, divided as defined in \S1
into metal-rich and metal-poor regimes at [Fe/H] = $-$1, populate two distinct 
regimes in correlations with phase, as shown in Figure~\ref{f3}.  
We have added curves to the figure to represent quadratic regression
lines for the two metallicity groups.
Model \teff's of metal-rich stars are approximately 250\,K hotter than 
those of metal-poor stars at all phases in the range 0.2 $< \phi <$ 0.6.  
Thus we confirm Skarka's (2014)\nocite{Skarka14} conclusion that metal-rich 
RRab stars are hotter than metal-poor RRab 
stars at a given $B-V$ color.\footnote{
Skarka's (2014)\nocite{Skarka14} mean $(B-V)_0$ colors are strongly 
correlated with those of \cite{Blanco92}. 
The latter were calculated for the phase interval 0.58 $< \phi<$ 0.80.
More recent studies, \eg, \cite{jur98}, use mean color definitions derived
from all parts of RR~Lyr pulsational cycles including the hotter phases near
$\phi$~= 0. 
These mean colors are, therefore, bluer than the intrnsic colors used 
by \citeauthor{Blanco92} to estimate interstellar reddening.}

The temperatures and gravities are correlated with 
metallicities, as illustrated in Figure\,\ref{f4}.
Solid black lines denote linear regressions of the quantities 
using the entire metallicity range.  
The horizontal dashed lines denote mean values of the metal-poor and
metal-rich stars.  
Standard deviations of these regressions for \teff\ and for \logg, 
listed in Table\,\ref{tab-sigmas} are indistinguishable, \ie, simple 
statistical analysis offers no guidance about choice of regressions.  
The physical bases for the natures of the correlations are uncertain.   
We defer further discussion of these two different statistical 
representations to \S5.

The most metal poor star in our sample, X\,Ari ([Fe/H]~=~$-$2.6), lies 
closer to the horizontal regressions for \teff\ and \logg.  
Data for additional very metal-poor stars collected and analyzed in a 
uniform way may help to resolve this uncertainty in choice of regressions.
We are aware of three such stars: NR\,Lyr, [Fe/H]~=~$-$2.54 {\citep{nemec13}};
CS 22881-029, [Fe/H]~=~$-$2.75; 
and CS~30317-056, [Fe/H]~=~$-$2.85 \citep{han11}.

\subsection{A Common Metallicity Scale for RRab Stars\label{allpapers}}

Spectroscopic metallicities have been derived for RR~Lyr stars beginning 
with the $\Delta S$ parameter introduced by \cite{preston59} to relate 
the strengths of the \species{Ca}{ii} K~line to the \species{H}{i} Balmer 
lines appearing on low-resolution spectra.
Many observations and calibrations of $\Delta S$ as a function of
[Fe/H] have been published over the last several decades.
The largest single source of K-line metallicities is the Lay94
the 302-star RRab survey.
Previous RR Lyr large-sample high-resolution spectroscopic metallicity and 
abundance ratio surveys include:
\citeauthor{clementini95} (1995, 10 stars); 
\citeauthor{lambert96} (1996, 18 stars);
\citeauthor{fernley96} (1996,  9 stars);
\citeauthor{for11b} (2011, 11 stars);
\citeauthor{liu13} (2013, 23 stars);
\citeauthor{nemec13} (2013, 41 stars); and
\citeauthor{pancino15} (2015, 18 stars).
We relate the metallicites reported in those studies to the 
[Fe/H] values derived here, using Layden's [Fe/H] values as the common
baseline.

In Figure~\ref{f5} we plot the differences between [Fe/H] values of various 
high-resolution studies (including our work) and those of Lay94}
as a function of the [Fe/H]$_{Lay94}$.
We have added simple linear regression lines to illustrate the mean
trends, which have positive slopes in five of the seven cases.
Inspection of this figure suggests that the most significant differences
among most high-resolution studies are metallicity scale offsets.

To attempt to merge together all of the high-resolution studies,
we first repeated the analyses of \cite{clementini95} and \cite{pancino15}, 
using only lines in common with our work.
We adopted their published $EW$s and our transition probabilities, and 
derived model parameters in the manner described in \S\ref{analysis}.
In general our results were not significantly different than theirs.

Then we shifted the [Fe/H] values of other studies onto our scale by
using the regression lines shown in Figure~\ref{f5}, and updated
regression fits to reflect our new analyses of \cite{clementini95} and 
\cite{pancino15} metallicities to calculate the offset between the
other studies and ours at a metallicity [Fe/H]$_{\rm Lay94}$~=~$-$1.25.
This metallicity point is of course arbitrary but is approximately in 
the middle of the observed RR~Lyr metallicity distribution.
In Figure~\ref{f6} we redraw Figure~\ref{f5} with the
re-analyses and applied [Fe/H] offsets.
Then all high-resolution analyses are merged in Figure~\ref{f7},
and we add an additional point to include the [Fe/H] value for the
eponymous star RR~Lyr from the analysis by \cite{kolenberg10}.
There is general agreement on the relationship between high-resolution
spectroscopic metallicities and the Lay94 values, and reasonably
small scatter in this relationship..
We stress that this exercise is not intended to be advocacy for 
our own RR~Lyr metallicity scale.
It simply suggests that with small scale shifts the [Fe/H] values of 
major RR~Lyr high-resolution studies can be brought into good agreement.

Two stars deserve special comment because they help anchor the 
high and low ends of the RR~Lyr metallicity scale.
As such they have been well represented in past and present analyses.
At the very metal-poor limit there is X~Ari, with 
[Fe/H]$_{\rm Lay94}$~=~$-$2.40.
There is some discord among the high-resolution results:
[Fe/H]~=~$-$2.50 (\citealt{clementini95}, or $-$2.51 in our reanalysis);
$-$2.47 \citep{lambert96};
$-$2.74 \citep{nemec13};
$-$2.14 (\citealt{pancino15}, or $-$2.20 in our reanalysis); and
$-$2.60, this study.
From all five studies, $\langle$[Fe/H]$\rangle$~=~$-$2.49 with $\sigma$~=~0.22.
Application of the scale offsets described above does not significantly
shrink the total envelope of values, as can be seen in the lower panel
of Figure~\ref{f7}.
The revised high-resolution mean value is $\langle$[Fe/H]$\rangle$~=~$-$2.61 
with $\sigma$~=~0.25.
The shift from [Fe/H]$_{\rm Lay94}$ for X~Ari is in agreement with the general
trend line of our analysis, but renewed attention to this star would be
welcome.

SW~And is one of the most metal-rich RR Lyr stars with
[Fe/H]$_{\rm Lay94}$~=~$-$0.38.
Most high-resolution studies report higher metallicities for this star,
but with scatter: 
[Fe/H] =~$-$0.06 (\citealt{clementini95}, or $-$0.24 in our reanalysis);
$-$0.26 \citep{fernley96};
$-$0.32 \citep{lambert96};
$-$0.07 \citep{liu13}; and
$+$0.20 \citep{nemec13}.
The means before and affer re-analyses and offsets are
$\langle$[Fe/H]$\rangle$~=~$-$0.10, $\sigma$~=~0.20, and
$\langle$[Fe/H]$\rangle$~=~$-$0.27, $\sigma$~=~0.14; for this star the
high-resolution analyses come into better accord with one another.  

With the scale shifts described above, a common relationship 
emerges between the high-resolution multi-star metallicity 
surveys considered here and the Lay94 K-line metallicities.
Our own data indicate that [Fe/H]$_{highres}$ = 
1.100$\times$[Fe/H]$_{\rm Lay94} + 0.055$ .
Future high-resolution studies can and should revisit this relationship,
which provides an important calibration for K-line surveys that now can
be efficiently conducted for large samples of Galactic and extragalactic
stellar systems.

\subsection{Relative Abundances of Other Elements\label{abratios}}

In \S\ref{analysis} we described the derivation of relative abundances
of several elements that can be studied even with moderate $S/N$ spectra
of metal-poor RR~Lyr stars; see Table~\ref{tab-abunds}.
Here we compare our [X/Fe] values to those reported by other abundance 
studies of RR~Lyr stars and by several surveys of non-variables over a 
wide metallicty range.
In Figure~\ref{f8} we plot [X/Fe] ratios as a function of metallicity
from: \textit{(b)} this study (black dots); from other RR~Lyr 
abundance studies (\citealt{clementini95}, \citealt{fernley96}, 
\citealt{lambert96}, \citealt{liu13}, and \citealt{pancino15}; red dots); 
and \textit{(c)} several large-sample studies of Galactic thin and thick 
disk main sequence and subgiant stars (\citealt{reddy03}, \citealt{reddy06}, 
\citealt{bensby14} green dots), and of Galactic halo low metallicity stars 
of various evolutionary states (\citealt{roederer14}; also green dots).
Relative abundance ratios are to first order insensitive to absolute
metallicity scale, so we go through the data in Figure~\ref{f8},
species by species.

In considering the data of Figure~\ref{f8}, note first that abundance 
ratios determined in our work are consistent with those reported in other 
RR~Lyr studies.  
Second, for most elements this agreement extends also to comparisons between
the RR~Lyr and non-variable stars.
Third, the RR~Lyr metal-poor abundance ratios are always in accord with those
of non-variable stars.
However, there are a couple of discordant cases to be noted among
the abundances of metal-rich stars, and so
here we will go through the figure on species-by-species basis.
\begin{itemize}

\item \textit{Ca, Figure~\ref{f8} panel (a):} 
All reported abundances are based on \species{Ca}{i} lines, which are plentiful 
(our values come from eight lines on average; Table~\ref{tab-abunds}).
The [Ca/Fe] ratios of RR~Lyr stars agree very well with those from other
stellar sample over the entire metallicity range.

\item \textit{Sc, Figure~\ref{f8} panel (b):} 
All reported abundances are based on a handful of \species{Sc}{ii} lines.
For halo stars the Sc abundances of RR~Lyr stars agree well with their
non-variable counterparts, but for disk stars the [Sc/Fe] values for RR~Lyr 
stars are 0.3$-$0.4 dex smaller than those of disk main sequence and
subgiant stars.  
One expects [Sc/Fe]~$\simeq$~0 at [Fe/H]~$\simeq$~0, so the RR~Lyr abundances
surely are wrong here.
Resolution of this problem, shared much more mildly by \species{Ti}{ii}
and \species{Ba}{ii} (see below), is beyond the scope of this paper.
Note that the first ionization potential of Sc is low ($IP$~=~6.56~eV),
ensuring that $N_{Sc II}$~$\gg$~$N_{Sc I}$, as is the second ionization 
potential (12.80~eV, lower than even that of hydrogen, 13.60~eV).
It is entirely possible that limitations of standard abundance analyses 
applied to the variable RR~Lyr stars are appearing in
the LTE calculations of Sc ionization.

\item \textit{Ti, Figure~\ref{f8} panel (c):} 
We have chosen to plot only abundances from \species{Ti}{ii} here, 
since in general this species has stronger transitions than does
\species{Ti}{i}.
There is a small Ti abundance offset, $\simeq$0.2~dex, between RR~Lyr and 
non-variable disk stars.

\item \textit{Cr, Figure~\ref{f8} panel (d):}
We were only able to measure a few lines of each Cr species, and so we
chose to average their abundance ratios for this figure.
Within the uncertainties of RR~Lyr analyses, there is agreement with
the non-variables at all metallicities.

\item \textit{Mn, Figure~\ref{f8} panel (e):}
The well-known sharply declining [Mn/Fe] values with decreasing metallicity
are reproduced well in RR~Lyr abundance surveys.
There are too-few points for stars with [Fe/H]~$<$~$-$2 to draw firm
conclusions on a possible negative offset relative to non-variable stars in 
this metallicity regime.

\item \textit{Ba, Figure~\ref{f8} panel (f):}
The agreement between RR~Lyr abundance ratios and those of non-variables
is satisfactory, given the large star-to-star scatter among all samples.
\end{itemize}

In summary, the abundance ratios derived for our RR~Lyr sample are mostly 
consistent with those seen in other stellar populations; further investigation
of Sc and Ba would be welcome.

\section{RADIAL VELOCITY AND H$\alpha$ FLUX MEASUREMENTS\label{rvhalpha}}

We accumulated 50-200 spectra for each program star.
These allow us to examine radial velocity variations of the metallic lines
and the H$\alpha$ line with pulsational phases, and to quantify the strengths 
of the emission component of the H$\alpha$ lines that occurs near phase
$\phi$~$\simeq$~0.9.
We did not consider the Blazhko stars BS~aps, UV~Oct, V1645~Sgr, and
AS~Vir here, leaving 24 stars for these analyses.
CD~Vel has been identified as a Blazhko variable \citep{szczygiel07} with 
$P_{Bl}$~= 66.25~d.
We included it in our analysis of stable RRab stars because its radial
velocity behavior appeared to be stable duing the entire time interval of our
observations ($\Delta HJD$~= 1070~d); see Figure~\ref{f9}.

Below we describe the derivation of velocities and emission measures.
For each star we show the results via three plots as a function
of pulsation phase, as displayed in Figure~\ref{f9}.
The left panel contains the radial velocities of the metallic lines
and H$\alpha$.
The velocity range shown is always 160~\kmsec\ with the boundaries shifted
appropriately for each star.
The middle panel has our H$\alpha$ emission measures, with the vertical
ranges chosen to clearly show emission for each star.
The right panel shows the ASAS $V$-band photometry, phased with
the period written in the panel legend.
The range in $V$ is always 1.6~magnitudes, shifted for display purposes.
Also in the legend is the stellar metallicity, computed as the mean
of [Fe/H] values derived from \species{Fe}{i} and \species{Fe}{ii} lines
(Table\ref{tab-abunds}).
We show in the manuscript only the data for X~Ari, and plots for
the other stars can be accessed in on-line files.

\subsection{Radial Velocities\label{rvmeasure}}

Thirteen orders of each RR\,Lyrae spectrum spanning the wavelength interval 
$\lambda\lambda$ 4000--4600\,\AA\ were flattened and stitched together by 
use of an IRAF script prepared by Ian Thompson (private communication).  
This is the same spectral interval used in the radial velocity 
investigations of \cite{preston00,chadid13}. 
Radial velocities of metal lines in these spectra were obtained by use of 
the IRAF/fxcor package.  
Masks centered on H$\delta$ and H$\gamma$ were employed to avoid deleterious 
broadening of the cross--correlation function by these wide, strong lines.  
The reference spectrum for metal-lines is that of CS 22874--009 for which 
we adopted RV = $-$36.6~\kmsec, the value used by \cite{preston00}.  
The reference spectrum for H$\alpha$ is that of the 
metal-poor giant CS~22892--052 for which we adopted RV = +14.2~\kmsec. 
This particular value contributed to an appreciable systematic correction 
noted in the following paragraph.\footnote{
The tentative low-amplitude velocity variation of 
CS 22892--052 suggested by \cite{Preston01} has not been confirmed 
by additional observations and analyses.}
Outputs of the fxcor package include VHELIO, the radial velocity corrected 
for the earth’s orbital and diurnal motions, and HJD, the Heliocentric 
Julian Date needed to calculate pulsation phases.  

Numerous observations of HD140283, our primary radial velocity standard, 
for which RV = $-$170.9\,\kmsec \citep{Kollmeier13} were used to apply a 
systematic correction of $-$2.9 km/s to measured H$\alpha$ radial velocities.
Spectra of du Pont echelle order 53, which contains H$\alpha$ near its 
center, were prepared for measurement as follows.  
For each stellar spectrum we made an average spectrum of adjacent 
orders 52 and 54.  
We divided order 53 by a spline fit to the continuum of this average 
spectrum, thus obtaining a flattened normalized spectrum of order 53 
in which fluxes at H$\alpha$ can be measured in units of the continuum 
flux at 6560\,\AA. 
These spectra were also used to measure the radial velocity of H$\alpha$. 

We used the Gaussian fitting function of IRAF/fxcor package to locate the 
centers of H$\alpha$ absorption features.  
In the course of our measurements we discovered that the ``Doppler'' 
core of the H$\alpha$ line is noticeably asymmetric during substantial 
fractions of pulsation cycles, the slope of one wing (sometimes red, 
sometimes violet) being greater than the other.  
In all such cases we fit a Gaussian to the flux minimum and the 
\textit{steepest} wing of the profile, arbitrarily attributing the other 
broader wing to unspecified asymmetry in line-of-sight gas motion, perhaps due 
to weak shocks at some pulsation phases. 
We don't $know$ why the profiles are asymmetric, but we needed a rule in order 
to measure them in a consistent manner.  
For a description of the shock complexity that accompanies RR~Lyrae 
pulsation see the observational evidence presented by \cite{chadid13}.   
The magnitude of effects resulting from use/non-use of our measurement 
procedure is almost always less than 2\,\kmsec, and usually less than 
1\,\kmsec, metrics that characterizes the inherent uncertainty of 
H$\alpha$ radial velocity estimates made with the du Pont echelle 
spectrograph.

\subsection{H$\alpha$ Emission Flux Measurements }

Emission flux measurements were made with the IRAF/splot package as sums 
of pixel counts in the H$\alpha$ profile above a reference continuum.  
When the emission profile lies entirely above the continuum, the measurement 
is unambiguous.  
However, when the emission is flanked by violet and red absorption 
components, an inescapable ambiguity arises about the portions of the flux 
between the absorption components that are due to overlapping absorption 
wings and those due to bona fide line emission.  
In instances when the flux maximum between absorption components lies below 
the continuum level, skeptics may deny the existence of $any$ emission.  
Justification for emission measurements in such cases rests on an appeal 
to phase continuity; namely, it would be unreasonable to suppose that 
emission ceased exactly at the moment when the apparent emission peak 
fell below continuum level.  
On the other hand, deciding exactly when to discontinue emission measurements 
and to begin measurement of violet-shifted absorption velocities requires a 
subjective judgment that we base on apparent radial velocity.  
When a violet absorption minimum first appears, its apparent velocity is 
strongly negative relative to the velocity measured minutes later when 
the violet and red minima have equal depths.  
Examples of rapidly increasing sequences of such 
``apparent velocities'' near phase 0.9 can be seen in left panels of 
a number of stars in Figure~\ref{f9}.
We deliberately included such (spurious) measures in the RV plots 
to illustrate how we decided when to treat the apparent violet minima 
as valid indicators of mass motion in the atmosphere. 
In our view the proper moment occurs when a cusp in the expanding 
velocities occurs near mid--rising light.  
Prior to that moment we regard the violet-displaced flux minimum in the profile 
merely as an $apparent$ absorption minimum located between the violet 
Stark absorption wing and the violet wing of approximately undisplaced 
central emission.

During primary light rise Balmer line emission (different in each star 
and overlain by red--displaced absorption) appears briefly in many 
RRab spectra, as illustrated in Figure~6 of \cite{preston64}.  
We use the emission flux in the violet wing of the H$\alpha$ profile as an 
admittedly imperfect proxy for the total (un-measurable) flux.  
Measurements of this H$\alpha$ flux were made as follows.  
A reference continuum was chosen at the violet--displaced flux minimum 
in the H$\alpha$ profile, as illustrated by the horizontal red line in 
Figure\,\ref{f10}.  
Uncertainties of unknown magnitude associated with this choice of 
reference level undoubtedly occur from exposure to exposure during 
primary light rise because of rapidly changing contributions from two 
absorption profiles to the total flux.  
Flux counts in the emission profile above the reference level (the red 
horizontal line in Figure\,\ref{f10}), are expressed in units of the 
local continuum fluxes at the phases of observation.  
To obtain true relative fluxes it would be necessary to multiply the 
plotted fluxes by the phase-dependent continuum fluxes at 6560\,\AA, 
which increase rapidly during the H$\alpha$ emission phase sequences.

Table\,\ref{tab-flux} contains a summary of quantities derived from the 
radial velocity and flux measurements.  
For five stars we list two metal velocity amplitudes that differ by as 
much as 4~\kmsec\ in different cycles.  
Our survey data do not provide enough information to determine whether 
these velocity differences are due to low-amplitude Blazhko modulation or 
to ``irregularity'' of the sort reported by \cite{chadid00}.  
For consistency we use the largest of these values in all graphical 
displays that follow.

\subsection{Photospheric motion parameters}

From a du Pont echelle heliocentric radial velocity $V_r(t)$, we derive a 
pulsational velocity $\dot{R}(t)$ in the stellar rest frame using:
\begin{equation}
\dot{R}(t) = -p\,(V_r(t) - V_{*})
\end{equation}
where $p$ is the value of the correction factor for geometrical projection 
and limb darkening and $V_{*}$ the center-of-mass velocity.
Because we derive radial velocities from metallic lines formed near the 
photosphere, the pulsational velocity $\dot{R}(t)$ is approximately that 
of the photosphere.
We adopt in this study the projection facto $p$ = 1.36.
Use of this approximation is discussed by \cite{chadid13}. 
Using polynomial fitting, we calculate the $\gamma$-velocity as the average
value of the heliocentric radial velocity curve over one pulsation period. 
The derivative of $\dot{R}(t)$ gives the dynamical acceleration $\ddot{R}(t)$ 
of the stellar surface layer in which metallic absorption lines are formed:
\begin{equation}
\ddot{R}(t) = d\dot{R}(t) / dt
\end{equation}
Performing the integration of $\dot{R}(t)$, we derive the photospheric radius 
variation $\Delta R(t)$:
\begin{equation}
\Delta R(t) = R(t) - R_{ph}=\int^{t}_{0} \dot{R}(t) dt
\end{equation}
where $R_{ph}$ is the mean photospheric radius of the star.

Radius and acceleration curves are respectively computed from integration and 
derivative of radial velocity curve. 
During the pulsation period  (0 $<$ $\phi$ $<$ 1) the gaps in the measurement 
data have been interpolated linearly, and, in order to minimize the noise, 
the raw radial velocitie curves have been convoluted by a sliding window of  
$\Delta$$\phi$~=~0.1 width.

Table\,\ref{tab-rv} contains a summary of photospheric 
motion parameters derived from the photospheric heliocentric radial 
velocity of the metal-poor and metal-rich RR\,ab stars.
The acceleration and the radius variation curves for individual 
metal-poor and metal-rich RR\,ab stars, as well as tha means for the 
two groups, are shown in Figure\,\ref{f11}. 
The acceleration curves exhibit two significant peaks. 
An acceleration peak (hereafter primary acceleration) occurs near phase 
$\phi$~=~0.90~$\pm$~0.02; it corresponds to the phase of minimum radius. 
A \textit{much} smaller secondary acceleration peak occurs near phase 
$\phi$ = 0.72~$\pm$~0.02. 
These secondary accelerations are clearly visible for several of the 
stable RRab stars in Figure~5 of \cite{chadid13}.   
Any acceleration associated with the $lump$\footnote{
The terms $hump$, $lump$, $rump$, and $jump$ follow the nomenclature 
of \cite{chadid14}}, 
$rump$, and $jump$ in the RV curve \citep{chadid14} that may be present 
around $\phi$~=~0.30, $\phi$~=~0.10 and $\phi$~=~0.70 respectively 
are at or below the level of detection.

\cite{chadid13} call primary acceleration the dynamical gravity of the star. 
We derive the average of the gravitational acceleration,
\begin{equation}
g∗ = G \frac{M}{R^{2}} 
\end {equation}

\noindent where $G$ is the gravitational constant 
($G\,=\,6.67\,\times\,10^{-11}$), and
$M$ and $R$ are respectively the mass and radius of the star. 

Figure\,\ref{f11} shows that almost all metal-poor RRab stars have 
radius variations larger and primary accelerations smaller than those 
of metal-rich RRab stars. 
Mean radii of metal-poor and metal-rich RRab stars occur at very nearly the 
same phases, $\phi$~= 0.65~$\pm$~0.02 during infall and 
$\phi$~=~0.12 $\pm$ 0.02 during outflow.
The amplitude of radius variation is larger by 50\,\%  in the metal-poor as 
it is in the metal-rich RRab stars.
The radius variations indicate that RRab stars are relatively extended stars 
for about half of their pulsation cycles; the minimum radius occurs near 
$\phi$~=~0.92~$\pm$~0.02 and the maximum value at 
$\phi$~=~0.38~$\pm$~0.02.  

\cite{de10} claim to have detected a rapid collapse phenomenon in the 
atmosphere of RR\,ab stars by the use of multichannel Str{\" o}mgren 
photometry. 
They suggest that at phase $\phi$~=~0.90 the atmosphere begins to 
collapse, quickly reaching a high gas density; then within a small phase 
interval $\Delta$$\phi$~=~0.10, the atmosphere expands again. 
No rapid collapse phenomena associated with the radius variation as suggested 
by \cite{de10} are  detected in our spectroscopic data.

The physical origin of the primary acceleration, the dynamical gravity of 
the star, is due to the $Sh_{H+He}$, the main shock produced by the 
$\kappa$ and $\gamma$ mechanisms, processes based on the effects of 
radiative opacity in $He^{+}$  and $He^{++}$ ionization zones that drive the 
pulsational instability in RR\,ab stars \citep{Cox80}.
The secondary acceleration is caused by the shock $Sh_{PM3}$, described 
in \cite{chadid14}, that produces compression heating. 
We suggest that primary acceleration is a better proxy for the intensity 
of the $Sh_{H+He}$  main shock. 
Accordingly, we use rise time, the very short time 
interval of the descending branch of the radial velocity curve to 
characterize the shock intensity.

Nearly all RR\,ab stars of this study possess more or less prominent 
secondary radial velocity maxima, the elbow near phase 
$\phi$~= 0.72~$\pm$~0.02. 
The phase and visibility of this maximum vary from star to star. 
It occurs late for WY\,Ant, and is indistinct for DT\,Hya and Z\,Mic. 
The secondary acceleration is largest for the metal-rich star HH\,Pup, 
which also exhibits the largest dynamical gravity. 
Metal-poor star VY\,Ser has the lowest dynamical gravity in our sample. 

We collect all of the derived motion parameters in Figure~\ref{f12},
using the primary acceleration (dynamical gravity) as the common variable 
on the figure's horizontal axis.
The various correlations seen in Figure~\ref{f12} will be discussed in
subsequent sections.
Inspection of this figure show that sevaral motion parameters 
have fundamentally different relationships with primary acceleration,
such as radius variation (panel b), duration of H$\alpha$ doubling $\Delta$t 
(panel d, discussed in \S7.1), and H$\alpha$ velocity amplitude 
$\Delta$RV$_{\rm H\alpha}$ (panel e).
In contrast, the other motion parameters show a single trend but different
(albeit overlapping) regions occupied by the metal-poor and metal-rich
RR~Lyr stars.

\section{EMISSION LINES AND SHOCK WAVES}

Shock pulsation models {\citep{Zel66}} suggest that the shock structure 
can be subdivided into five principal zones with different physical 
characteristics: the precursor zone, the shock front, the thermalization zone, 
the ionization zone, and the zone of the cooling wake which is partially 
transparent in the Balmer and weak spectral lines. 
The wake zone, just behind the shock front, is a relatively narrow region 
in comparison with the photospheric radius \citep{chadid11}. 
The evolution of RR~Lyrae $H\alpha$ line profiles gives crucial information 
about nonlinear dynamics in the extended atmosphere above the wake zone.
The wings of $H\alpha$ lines are formed deep in the atmosphere while 
the line cores are formed higher. 
We study the Doppler effect in these cores.

\cite{chadid14} report photometric evidence for multiple shock waves crossing 
the  envelope of the RR\,ab Blazhko star S~Arae.  
These shocks, $Sh_{PM1}$, $Sh_{PM}$, $Sh_{PM2}$, $Sh_{PM3}$, and $Sh_{H+He}$ 
appear respectively during the $jump$, $lump$, $rump$, $bump$ (secondary 
acceleration) and $hump$ (primary acceleration) in the light curve, with 
different amplitude and physical origin.
When shocks traverse the envelope, they heat the sub-photospheric 
gas that caused a local decrease of the opacity and consequently they 
induce a luminosity increase in the light curve.
\cite{chadid14} proposed a new shock propagation scenario and showed that 
during the time interval of the ascending branch of the radial velocity 
curve all shocks are receding in Eulerian coordinates system while they 
are advancing shocks in the Lagrangian coordinates system.  
The brightest $H\alpha$ emissions observed during the $hump$ and $bump$ 
are directly produced in the radiative wake of respectively 
the $Sh_{H+He}$ main shock caused by the $\kappa$ and $\gamma$ mechanisms,
while the $Sh_{PM3}$ is caused mainly by collisions and compression waves.  
The  $H\alpha$ absorption lines show doubling of their cores
 during the $hump$ (primary acceleration). 
According to the classical Schwarzchild doubling effect 
\citep{schwarzschild52}, the phenomenon can be explained by the propagation
of the main shock, $Sh_{H+He}$. 
No doubling of the $H\alpha$ absorption lines has been detected during the 
$bump$ in the light curve heretofore.

\section{FIRST DETECTION OF H$\alpha$ LINE-DOUBLING DURING THE BUMP\label{shock}}

We have obtained clear evidence of H$\alpha$ line-doubling during the 
photometric $bump$ associated with the secondary acceleration
in two of the brightest RRab stars in our sample, WY Ant and RV Oct. 
It begins near the phase of violet-displaced H$\alpha$ emission reported
by \cite{gillet88}.
We use the spectra of RV Oct here for illustrative purposes. 
Marginally resolved H$\alpha$ line--doubling during the phase interval 
0.63$-$0.85 is  common, if not ubiquitous, among the the metal-poor RRab 
stars of our sample.
It occurs during the inward motion of a shock in the Eulerian coordinates 
system.

\subsection{Apparent ballistic acceleration and deceleration  \label{aacc}}

We suspect that the violet emission edge of the H$\alpha$ that first 
appears near phase 
$\phi$~= 0.62~$\pm$~0.02 in the top panel of Figure\,\ref{f13} is 
responsible for the small, apparent increase in radial velocity seen in the 
bottom panel of Figure\,\ref{f13} at this phase. 
This emission pushes measured radial velocities to spuriously large values, 
enhancing the brief secondary velocity maximum. 
The apparent deceleration, the $elbow$\footnote{
The term elbow follows the nomenclature of \cite{chadid13}}, 
that follows is due to heretofore unresolved line-doubling during the 
$Sh_{PM3}$ shock. 
A line component of longer wavelength produced by ballistic infalling gas 
also appears suddenly at phase 0.63 and strengthens as the 
original line weakens and gradually disappears. 
Because IRAF/fxcor does not clearly resolve this line-doubling in most of our 
spectra, we usually do not detect the velocity discontinuity depicted 
schematically by the blue lines in the bottom panel of Figure\,\ref{f13}. 
Note that both line components are formed by infalling gas at the interface of 
the  $Sh_{PM3}$  ballistic shock. 
Such doubled profiles are not present in any of our spectra of metal-rich 
RRab stars.

\subsection{Bump line-doubling mechanism\label{dblebump}}

The $Sh_{PM3}$ ballistic shock {\citep{chadid14} crosses the inward moving 
atmosphere during the phase interval phase $\Delta$$\phi$ [0.63,0.85]. 
H$\alpha$ line emission is produced in the radiative wake of the ballistic 
shock.
Figure\,\ref{f14} illustrates schematically the sequence of $H\alpha$ 
profiles during the phase interval 0.62$-$0.76.  
The $H\alpha$ profile shows first a redshifted absorption line 
(hereafter original line).
At phase $\phi$~=~0.62 a blueshifted emission appears while the velocity of 
the original line increases and its residual intensity decreases. 
A new second redshifted absorption line appears at phase $\phi$~=~0.63 with 
a velocity greater and a residual intensity smaller than the original line. 
At phase $\phi$~=~0.66, the residual intensity of the second redshifted line 
increases to be equal to that of the original line. 
At  phase $\phi$~=~0.73, the second  redshifted line strengthens as 
the original line weakens and gradually disappears. 

We interpret these variations in the $H\alpha$ profile 
during the bump in the light curve as follows.
We assume that the $Sh_{PM3}$ ballistic shock crossing the infalling 
$H\alpha$ line forming zone has an inward motion in the Eulerian rest frame 
although is moving outward in the mass Lagrangian coordinates system, 
and that its existence is strongly dependent on the amplitude of 
the ballistic motion.
Initially the velocity of $Sh_{PM3}$ is not large enough to produce 
resolved doubling.
The shock, moving outward through mass shells, is accelerated during its 
propagation; consequently a second redshifted absorption component is 
observed with a higher velocity than the original line. 
Its residual intensity increases as the original line weakens at phase  
$\phi$~=~0.66.  
At this time the $H\alpha$ doubling is clearly visible and the shock 
amplitude is near 50~\kmsec. 
The speed of sound is almost entirely dependent on its temperature, 
being roughly equal to 0.11 $\times$ $\surd$ $T_{eff}$ in ~\kmsec\ units.  
Adopting T $\sim$ $T_{eff}$ = 6050\,K for RV Oct (Table~\ref{tab-models}), 
the Mach number of the the shock $Sh_{PM3}$ is approximately 8.

Thus a shock Mach number around 8 is necessary to produce the $H\alpha$ 
line-doubling phenomenon. 
Later at phases 0.76--0.85, the original line gradually disappears. 
At phase $\phi$~=~0.85, the shock leaves the  $H\alpha$ forming zone as the
star approaches the phase of the minimum radius.  
Soon thereafter, the shock $Sh_{H+He}$, associated to the primary 
acceleration, emerges from the photosphere and reverses
the motion of infalling atmospheric layers.

The weak H$\alpha$ emission, associated with the inward shock $Sh_{PM3}$, 
appears blue--shifted, which is inconsistent with what we expect to see,
namely red-shifted emission. 
This is puzzling. 
We conclude that more than one mechanism must be at work to produce
the observed profile.
We suggest that an atomic mechanism, some combination of Stark and Doppler
broadening, and a geometric mechanism $-$ some unanticipated motions of
gas layers $-$ may be responsible.
We simply do not know how these mechanisms combine to produce what is observed.

In closing this section we call attention to the visibility of the elbow 
in the metal-RV curves of Figure~\ref{f9}.
The elbow is well-marked, or at least identifiable in most metal-poor RRab 
stars.  
However, it is not present in metal-rich ST Vir, DX Del, V445 Oph, AV Peg, and 
AN Ser, and it is only marginally visible in large-amplitude HH Pup and W Crt.
We shall see in \S7 that this distinction is related to the 
different atmospheric extents of metal-poor and metal-rich atmospheres.

\section{COMPARISON OF DYNAMICS IN METAL-RICH AND METAL-POOR RRab ENVELOPES}

\subsection{Difference in upper--atmosphere structure}

The strong albeit imperfect correlation of V-light amplitude with metallic 
line RV amplitude in Figure~\ref{f15} shows, first of all, that these two 
amplitudes are equally acceptable indicators of RRab pulsation amplitude. 
Residuals of individual stars in these regressions can be understood as a
combination of errors in the amplitude estimates and small (non-Blazhko) 
irregularities in pulsation amplitudes in some stars \citep{chadid00}. 
In particular, the rms scatters of ASAS $V$ magnitudes that produce the $V$ 
light amplitudes in Table~\ref{tab-flux} are strongly correlated with apparent 
magnitude because signal-to-noise decreases with decreasing luminosity 
in the ASAS program.
The well-known period $versus$ light amplitude relations among RRab stars 
\citep{clement99} and their associated period $versus$ velocity amplitude 
relations undoubtedly contribute to the scatter as well.  
Investigations of the latter are beyond the scope of this paper. 
Slopes of the regressions calculated separately for metal-rich (red) and 
metal-poor (blue) in this diagram differ by less than the sum of their 
standard errors, regardless of the choice of independent variable in the 
least squares solutions. 
Thus metal-poor and metal-rich RRab stars share a common relation between 
luminosity amplitude and photospheric velocity amplitude. 

However, the $H\alpha$ lines of metal-poor and metal-rich stars which are
formed at very low optical depths (\S5) behave very differently in 
metal-poor and metal-rich stars, as illustrated in Figure~\ref{f16}. 
The $H\alpha$ velocity amplitudes of the metal-rich stars are systematically 
smaller by $\sim$20~\kmsec\ than those of metal-poor stars at a given 
metal velocity amplitude. 
The velocity structures and geometrical extents of the outer atmospheres of 
the two groups must be very different.

The durations of double  $H\alpha$ lines in the two abundance groups shown 
in Figure~\ref{f17} provide an additional indication of different 
atmospheric structures. 
We define duration as the time interval $\Delta$t during which IRAF/splot 
can identify two absorption minima in the H$\alpha$ profile.
The durations are markedly longer in stars with large H$\alpha$ emission fluxes.
We note that three of the four stars with $no$ observed $H\alpha$ emission 
flux (AV\,Peg, V445\,Oph, and W\,Crt) have the very short periods that 
attracted Kukarkin's (1949)\nocite{kukb49} attention long ago.
The fourth, DX\,Del (P~=~0.473d), has the lowest light and RV amplitudes 
in our metal-rich sample.

The ranges in appearance and duration of line-doubling is illustrated in 
the montages of $H\alpha$ profiles for metal-rich AV~Peg and metal-poor 
RV Oct in Figure~\ref{f18}. 
The profiles of AV Peg are typical of the four stars mentioned above,
i.e. they are devoid of $H\alpha$ emission in our du Pont echelle spectra. 
A violet absorption wing turns into a resolved absorption feature that 
replaces the red component on the short time scales plotted in 
Figure~\ref{f17}. 
In AV Peg line doubling can be seen only in the brief interval from 
$\phi$~=~0.93 to $\phi$~=~0.96, $\Delta$t = 0.03\,P = 1000\,s, 
corresponding to 3\,\% of pulsation period. 
The same datum for RV\,Oct is $\Delta$t~=~0.17\,P\,=\,8400\,s, 
corresponding to 17\,\% of pulsation period; the durations differ by a 
factor $\sim$\,8. 
These durations are measures of the times that out-flowing gas sweeps up 
and reverses the flow of in-falling gas.
They define the lifetime of the $Sh_{H+He}$ main shock in the upper atmosphere.
The lifetimes of the main shocks in metal-poor and metal-rich stars 
differ by a factor $\sim$8.
For fluxes less than 1 continuum unit, the metal-poor durations exceed 
metal-rich durations by a factor of two or less.
Note, however, that durations as large as 0.12\,d $\sim$\,10000\,s occur 
among the metal-poor stars plotted in Figure~\ref{f17}.

\subsection{Difference in photospheric structure}

The panel (d) of Figure\,\ref{f12} tells us that large photospheric 
dynamical gravities (i.e., strong shocks) in the photospheres of metal-rich 
stars are of short duration, while weaker shocks in metal-poor stars are of 
longer duration. 
We make a dynamical gravity division between the metal-poor and metal-rich 
families near to $\ddot{R}$=30\,$km s^{-2}$. 
To this, we add the well-marked correlation between $H\alpha$ flux and duration 
of the weaker shocks of metal--poor stars shown in Figure~\ref{f17}.
Finally, recall that the photospheric radius variation is larger for the 
metal-poor RRab stars than for the metal-rich stars (bottom panel of 
Figure~\ref{f11}); the average difference in the panel (b) of 
Figure~\ref{f12} is approximately $\Delta R$~= 0.17\,R$_{\bigodot}$.

All the regressions in Figure~\ref{f12} indicate that dynamical 
gravity, the primary acceleration, is the empirical, unifying, causal 
parameter of the various observed envelope phenomena.  
Used as independent variable it produces simple linear regressions for 
secondary acceleration, radius variation, rise-time, duration of $H\alpha$ 
doubling, and hydrogen and metallic radial velocity amplitudes. 
The slopes and intercepts of the metal-rich regressions (red-straight lines) 
are different from those of their metal--poor counterparts (blue-straight 
lines). 
The differences between the metal-rich linear regressions and those of their 
metal--poor counterparts are more pronounced in the upper-atmospheric 
parameters (duration of $H\alpha$ doubling, $H\alpha$ radial velocity 
amplitude and $H\alpha$ emission) than those of the photosphere (secondary 
acceleration, metal radial velocity amplitude and rise--time). 
Finally, the radius variation is larger by a factor $\sim$\,1.8 for metal-poor 
stars than that of their metal--rich counterparts.

To conclude, the motions of the upper-atmospheric and near-photospheric regions
exhivit very different behaviors, and these differences are more pronounced 
in metal-poor RRab stars than in their metal-rich counterparts.

\subsection{Interpretation}\label{interp}

Panel (a) of Figure~\ref{f12} shows the regression between the 
secondary and primary accelerations for metal-poor and metal-rich RRab: the 
stronger the dynamical gravity the more important is the secondary acceleration.
Both groups share a similar physical process of the amplification of the 
secondary acceleration with the primary acceleration which indicates dynamical 
coupling interplay between the $Sh_{H+He}$ main shock and  the $Sh_{PM3}$  
ballistic shock. 
The greater the dynamical gravity, the larger is $Mach$ number $Sh_{H+He}$ 
that causes higher level of dynamical imbalance, , i.e., the 
non-synchronization of motions in the upper and lower photospheric layers. 
Stronger collision induce higher supersonic motion, $Sh_{PM3}$. 
Thus secondary acceleration becomes more important. 
Two distinct correlations between metal-line (near-photospheric) radial 
velocity amplitude and primary acceleration exist for both 
metal-poor and metal-rich RRab stars (panel (f) of Figure~\ref{f12}). 
The amplitude of the photospheric ballistic motion is larger when the 
photosphere is crossed by a stronger shock, i.e., the more important primary 
acceleration, the larger is the metal radial velocity amplitude. 
The comparison of primary accelerations in the top panel of 
Figure\,\ref{f11} demonstrates that the intensity of the $Sh_{H+He}$ main 
shock is more important in the photospheres of metal-rich stars than 
those of metal--poor stars.

The most striking characteristic of the ballistic photospheric motion is its
greater magnitude in metal-poor relative to metal-rich RRab stars
(Figure~\ref{f12}).  
Longer period implies longer ballistic motion.
The greater extensions of the metal-poor envelopes complicates their 
atmospheric motions; the non-synchronization effect greatly changes atmospheric
motion structure and creates chaotic behavior, mainly in the upper atmospheres. 
Moreover, the region of formation of spectral lines is greatly enlarged with 
concomitant effects on line profiles.
This enlarged region of line formation in metal-poor stars is the origin 
of the greater duration of $H\alpha$ doubling (Figure~\ref{f12}, 
panel (d) ),
$H\alpha$ radial velocity amplitude (panel (e) of Figure~\ref{f12}) 
and the peak $H\alpha$ emission flux.

H$\alpha$ emission is generally weaker in metal-rich RRab stars 
(Figure~\ref{f19}).
$No$ emission could be detected in four of the eight metal-rich RRab stars
in our sample (AV\,Peg, V445\,Oph, ST\,Vir and DX\,Del), and the strongest
emission in the remaining four occurred in the lowest metallicity star 
of the metal-rich group (UU~Vir, [Fe/H]~= $-$0,93).
We attribute this metallicity effect to the fact that the metal-poor RRab 
atmospheres are more extended, and the amplitudes of the ballistic motions
of the metal-poor RRab atmospheres are greater than those of metal-rich RRab.

The bottom panel of Figure\,\ref{f19} shows that strongest H$\alpha$
emission in RRab stars is restricted to those with the largest hydrogen 
radial velocity amplitudes.
A threshold effect seems to be present: a steep gradient of $H\alpha$ 
emission occurs near H$\alpha$ radial velocity amplitude $\sim$\,115~\kmsec, 
when the duration of $H\alpha$ line-doubling is $\sim$\,10\,\% of the 
pulsation period.
The emission flux increases abruptly when the dynamical gravity exceeds  
$\ddot{R}$ $\sim$ \,20\,km~s$^{-2}$,   independent of the metallicity, 
pulsation period and radius variation. 
We call this phenomenon the hydrodynamic gadient.
This mechanism applies to metal--poor and perhaps to  metal-rich RR\,ab 
stars as well.  
Only one metal-rich RR\,ab star in our sample (UU~Vir, [Fe/H]~= $-$0,93)
shows this phenomenon, and it is the most metal-poor of the group.
The hydrodynamic gadient mechanism is strongly dependent of the temperature 
and the gravitational acceleration of the star, and affects the stars with 
lower temperatures ($<$ 6100\,K) and  slighter gravities ($log$$g<$ 1.9) 
(Fig~\ref{f4}).
Evidently the upper atmosphere reaches a high-hypersonic regime when the 
dynamical gravity exceeds $\ddot{R}$~$\sim$ 20\,km~s$^{-2}$.
Accordingly, we divide  the metal–poor RR ab stars into two types, type 1 
and type 2, and we define four main shock regimes in the upper atmospheres of 
RR\,Lyrae stars: \begin{enumerate}
\item 
Trans-sonic Regime: No hydrogen emission or line doubling is observed. 
This regime could characterize RRc and RRd.
\item
Supersonic Regime: Hydrogen line doubling and only a very small or even 
no hydrogene emission is observed. 
This regime occurs mostly in the metal-rich RRab.
\item
Hypersonic Regime: Additional He\,I emission line is observed. 
The temperature must be higher than 10\,000 K, with a threshold Mach number 
$Mach_{HeI}$. 
The hypersonic regime occurs in metal-poor RRab type\,1 and a few metal-rich 
RRab with higher dynamical gravity.
 \item
High-hypersonic Regime:  He\,II emission line is observed. 
The temperature and energy respectively higher than 30\,000 K and 24.59\,eV are
able to ionize a neutral helium atom with a threshold Mach number $Mach_{HeII}$.
The atmosphere is very extended and a circumstellar envelope around the star 
may occur. 
This should characterize the metal–poor RRab type 2 that could be cooler 
and more convective.
This regime has been already identified  in two metal--poor RR\,ab stars:
AS~Vir and V1645~Sgr \citep{preston11}.
\end{enumerate}

Finally, we consider briefly the H$\alpha$ emission of RRab stars in the 
context of other post-main sequence pulsating stars.  
We note the absence of Hα emission in most classical Cepheids with periods 
less than 10 days, i.e., those Cepheids with lowest luminosity and least 
extended atmospheres.  Among the remainder, the H$\alpha$ emission of the 
metal-poor RRab stars is weaker than those seen in W Virginis, RV Tauri, 
and Mira stars. 
The latter families comprise a sequence of increasing luminosity and 
atmospheric extent. 
Wider atmospheric extension in these cooler pulsators is accompanied by 
larger Mach numbers of Sh$_{\rm H+He}$. and stronger shocks.

 \section{ARE METAL-RICH AND METAL-POOR RRab TWO SEPARATE FAMILIES?}
 
We suppose that metal-rich and metal-poor RRab stars belong to a single 
family of helium-core-burning horizontal branch stars. 
However, our study reveals strikingly different characteristics
of these two subgroups.
Their atmospheric dynamics and structure are different. 
This difference is typically more pronounced in the upper atmosphere. 
The differences in their atmospheres are induced by large differences in
the mechanical energy of the ballistic motion and the strengths and locations 
of shock waves in the photosphere and upper atmospheric layers.
Indeed, according to \S5, the dynamical gravity and then 
the $Sh_{H+He}$ main shock intensity are larger in the photospheres of 
metal-rich RRab than they are in those of metal-poor RRab.
The $Sh_{H+He}$ main shock always propagates outwards and its amplitude 
increases with height as the density of the gas decreases until 
dissipation mechanisms kill the shock.
The $Sh_{H+He}$ main shock strengthens as it propagates highly into regions
of lower gas density and reaches high Mach numbers. 
The main shock intensity is larger in the upper atmospheres of metal-poor RRab 
than in those of metal-rich RRab.     

Our spectroscopic results show that the gravitational acceleration and 
effective temperature are smaller in metal-poor RRab than in metal-rich. 
From these considerations we hypothesize that: \\
(1) unlike the metal-poor stars, the metal-rich stars lie close 
to the blue edge of the fundamental $F$ mode instability region where the 
first overtone mode may disturb the fundamental mode by acting on the 
radiative loss of the $Sh_{H+He}$ main shock. 
In fact, from our results, the effective temperature is smaller in metal-poor 
than metal-rich RR\,ab stars. 
Accordingly, we hypothesize that the metal-rich stars are close to the 
blue edge of the fundamental instability strip. 
On the other hand, we might hypothesize that metal-rich RR ab stars are 
affected by a disregarded first overtone mode. 
The small amplitude of this latter is unable to invoke the double mode 
behavior similar to Bailey's RRd and has no sign in the frequency spectra.
This small first overtone excitation might generate a small perturbation that 
disturbs the $\kappa$ and $\gamma$ mechanisms process,
and therefore might react on the radiative loss of the main shock.
(2) The gravitational acceleration is smaller in metal--poor RR\,ab. 
Therefore the non-synchronization of their photospheric layers is
less pronounced.\\ 
(3) The metal-poor stars lie closer to the cooler fundamental red edge,  so
their convective envelopes are deeper. 
The thicknesses of their compression zones are greater, which may explain why
their photospheric radii and their radius variations are larger. 
The $Sh_{H+He}$ main shock starts crossing the metal-poor photosphere at a
lower optical depth with smaller $Mach$ number. \\
(4) {\cite{Huete10}}  finds that shocks encountering the density 
inhomogeneities characteristic of convective envelopes generate turbulent 
velocities that dissipate shocks. 
We attribute the smaller Mach number of the main shock in metal-poor 
photosphere to their larger convective envelopes.\\
(5) Metal-poor atmospheres are very extended, particularly those of 
metal-poor type\,2.  
In these stars the main shock propagates outwards into the upper extended 
atmosphere, and its amplitude increases with decreasing gas density, so it 
reaches higher Mach numbers in the upper atmosphere. 
A circumstellar envelope around the metal-poor RR\,ab type\,2 may occur,
perhaps as suggested recently by \cite{stellingwerf13}.
These issues will be addressed in a subsequent paper.

Finally, we interpret the hydrodynamic gradient in metal-poor RRab, an
abrupt hydrodynamic amplification process, as follows:   in addition to 
the $\kappa$ and $\gamma$ mechanisms, metal--poor RR\,ab type\,2 generate 
a third pulsation excitation mechanism, the radiation--modulated excitation 
mechanism, stronger than the $\kappa$ and $\gamma$ mechanisms that 
creates a new shock wave $Sh_{RME}$ inducing a stronger  main shock 
$Sh_{H+He}$ + $Sh_{RME}$ with higher Mach number and then an abrupt 
hydrodynamic amplification process. 
Indeed, such a pulsation excitation mechanism has been reported by 
{\cite{xiong98}} in the horizontal branch cooler convective envelopes 
with $T_{eff}$\,$<$6200\,K which are similar to the metal-poor RR\,ab type\,2 
(Sect.\ref{interp}) where the hydrodynamic gradient mechanism occurs. 
In fact, by the use of local time-dependant statistical theory of convection 
with a both the dynamic and thermodynamic coupling between convection and
oscillations, {\cite{xiong98}} demonstrates that the radiation-modulated 
excitation mechanism depends on the gradient of the radiative luminosity (see 
equation 14 in {\cite{xiong98}}) and takes place in a zone of the radiation 
flux gradient which is the  bottom and the top of the convective zone. 
This third process converts radiation energy into the kinetic energy of 
pulsation. 
We hypothesize that such radiation--modulated excitation could generate, in 
metal--poor RR\,ab type\,2, a shock wave, that we call $Sh_{RME}$.  
This latter interferes with the $\kappa$ and $\gamma$ mechanisms shock 
$Sh_{H+He}$ inducing a stronger resulting main shock ($Sh_{H+He}$ + $Sh_{RME}$) 
with an abrupt increase of  $Mach$ number and then the observed threshold 
effect in $H\alpha$ emission that occurs near H$\alpha$ radial velocity 
amplitude $\sim$\,115~\kmsec.
Further calculations are needed to clarify such synergies between 
hydrodynamic gradient and radiation--modulated excitation mechanisms.

\acknowledgments

We thank Jose Govea for initial analyses of some of the RR Lyr spectra.
We are happy to acknowledge our referee for helpful comments that improved the
manuscript.
We thank all the Las Campanas Observatory support personnel for their help 
during the course of our endeavor,  
We offer our special regards to several duPont telescope operators 
for their efforts in assisting with the observations required to produce 
this paper.  
Finally, we are most grateful to Stephen Shectman for inventing the duPont 
echelle spectrograph thirty-some years ago.
This work has been supported in part by NASA grant NNX10AN93G (J.E.L.), 
by NSF grant AST-1211055 (J.E.L.) and NSF grant AST-1211585 (C.S.).  




\clearpage
\begin{figure}
\epsscale{1.00}
\plotone{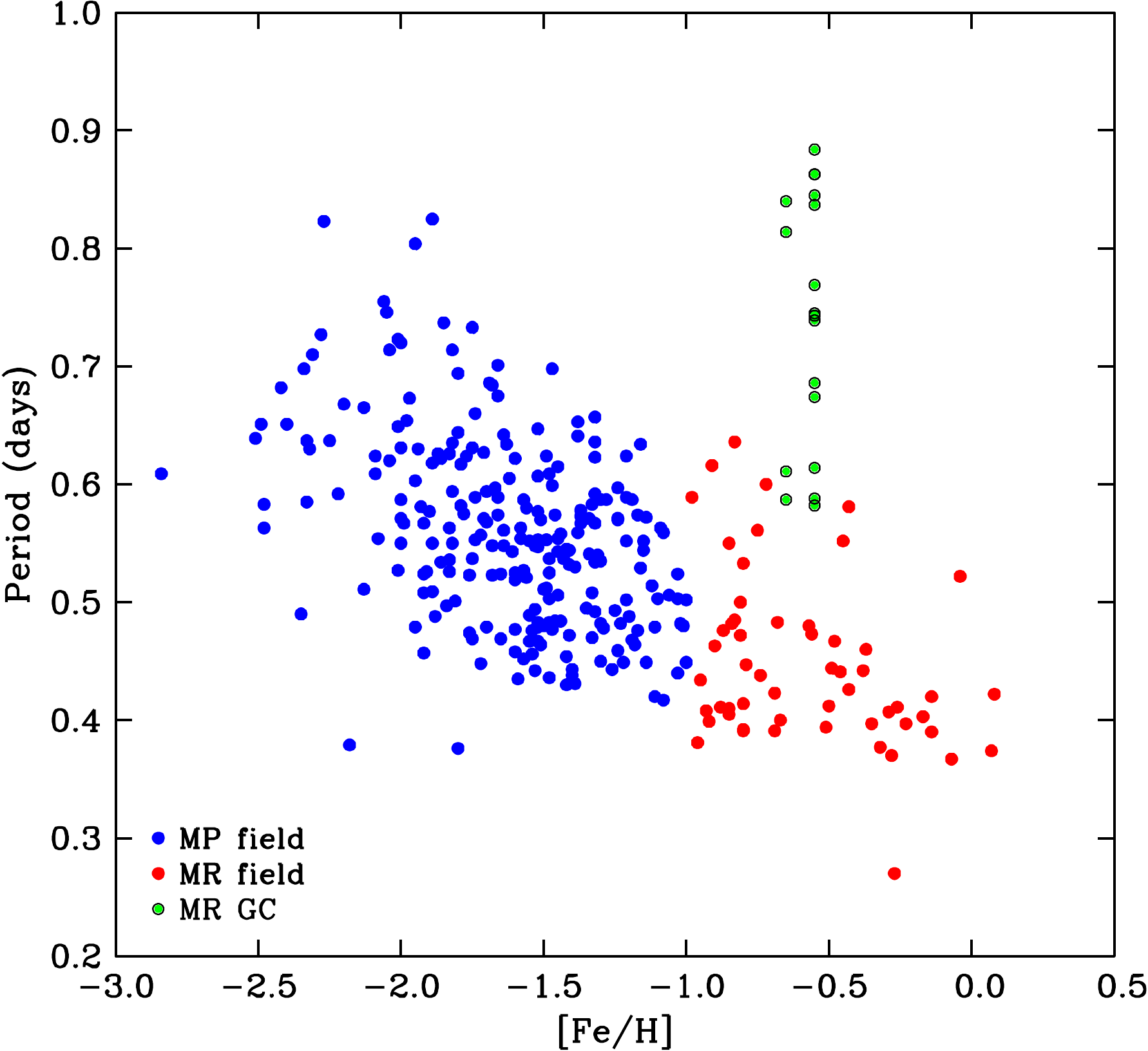}
\caption{\label{f1}
\footnotesize
The pulsation periods of the RRab stars of Layden(1994) are plotted 
$versus$ their [Fe/H]$_{\rm Lay94}$ values.
This figure is adapted from Figure~1 of \cite{layden95b}.}
As indicated in the figure legen, blue and red circles denote metal-poor 
(MP) and metal-rich (MR) stars, respectively.
This symbol convention will be followed in all subsequent figures that
divide stars into MP and MR categories.
Green circles denote the RRab stars in NGC 6388 and NGC 6441. 
\end{figure}

\clearpage
\begin{figure}
\epsscale{1.00}
\plotone{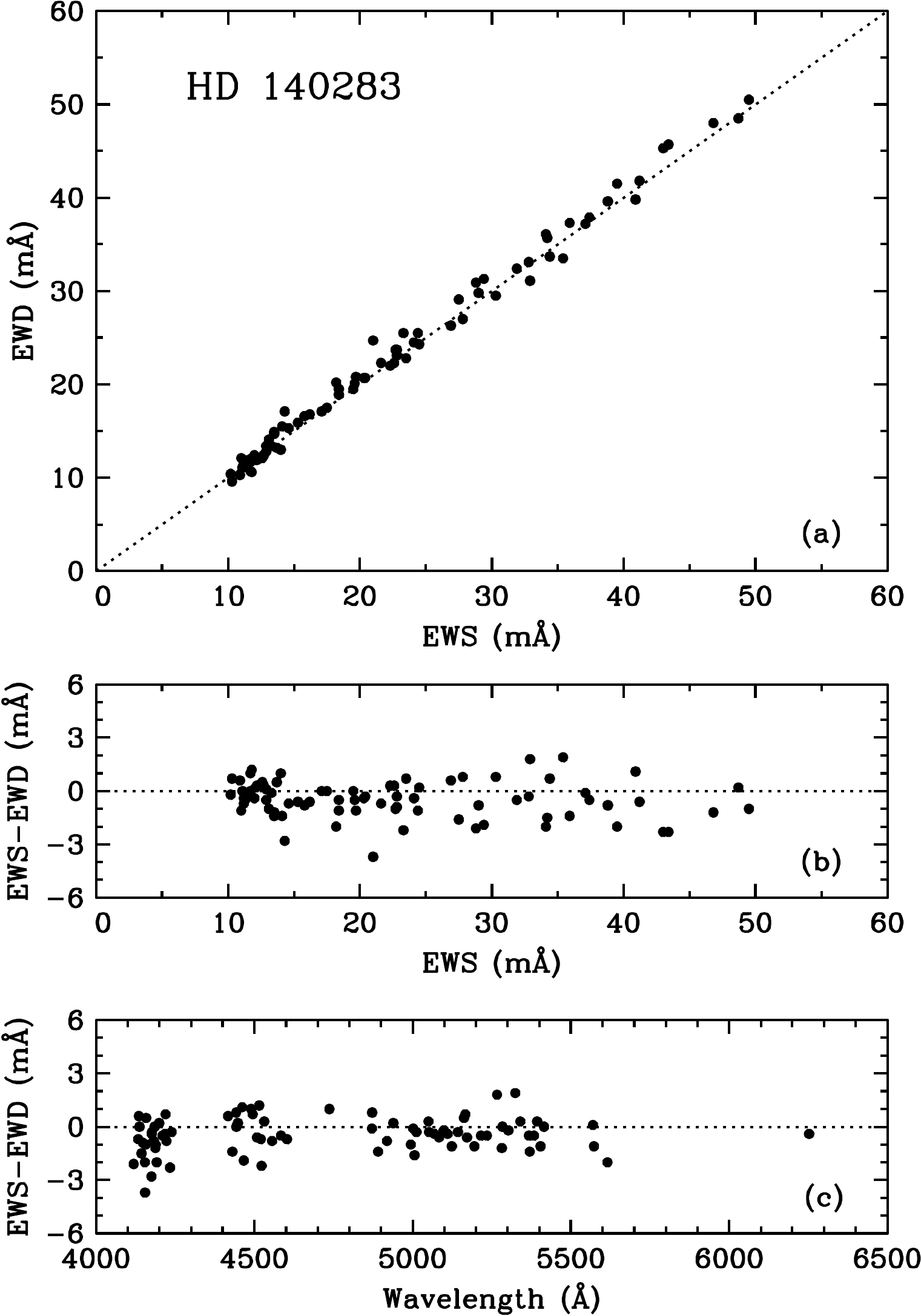}                                             
\caption{\label{f2}
\footnotesize
Equivalent widths HD\,140283 \species{Fe}{i} and \species{Fe}{ii} lines 
measured in du Pont spectra of (EWD) compared to the equivalent widths 
measured in a Suburu spectrum by \cite{Gallagher10} (EWS).
In panel (a) the two data sets are plotted against each other, and in
panels (b) and (c) the EW differences are plotted $versus$ EWS and
wavelength, respectively}
\end{figure}

\clearpage
\begin{figure}
\epsscale{1.10}
\plotone{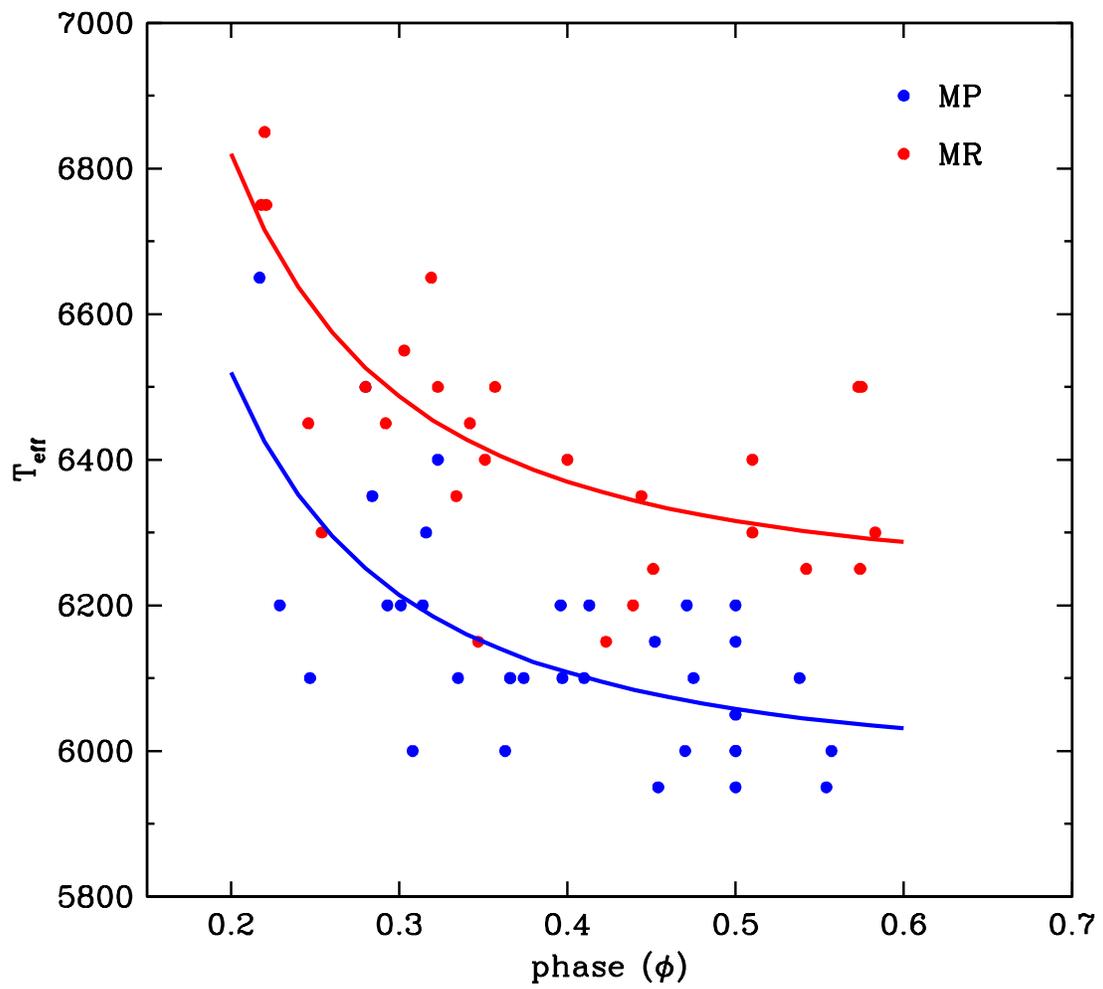}
\epsscale{1.00}
\caption{\label{f3}
\footnotesize
Correlations of derived \teff\ values with pulsational phases $\phi$ for
metal-poor and metal-rich stars.
The symbols are as in Figure~\ref{f1}.
The curves are computed regression lines to the two metallicity populations.}
\end{figure}

\clearpage
\begin{figure}
\epsscale{1.15}
\plotone{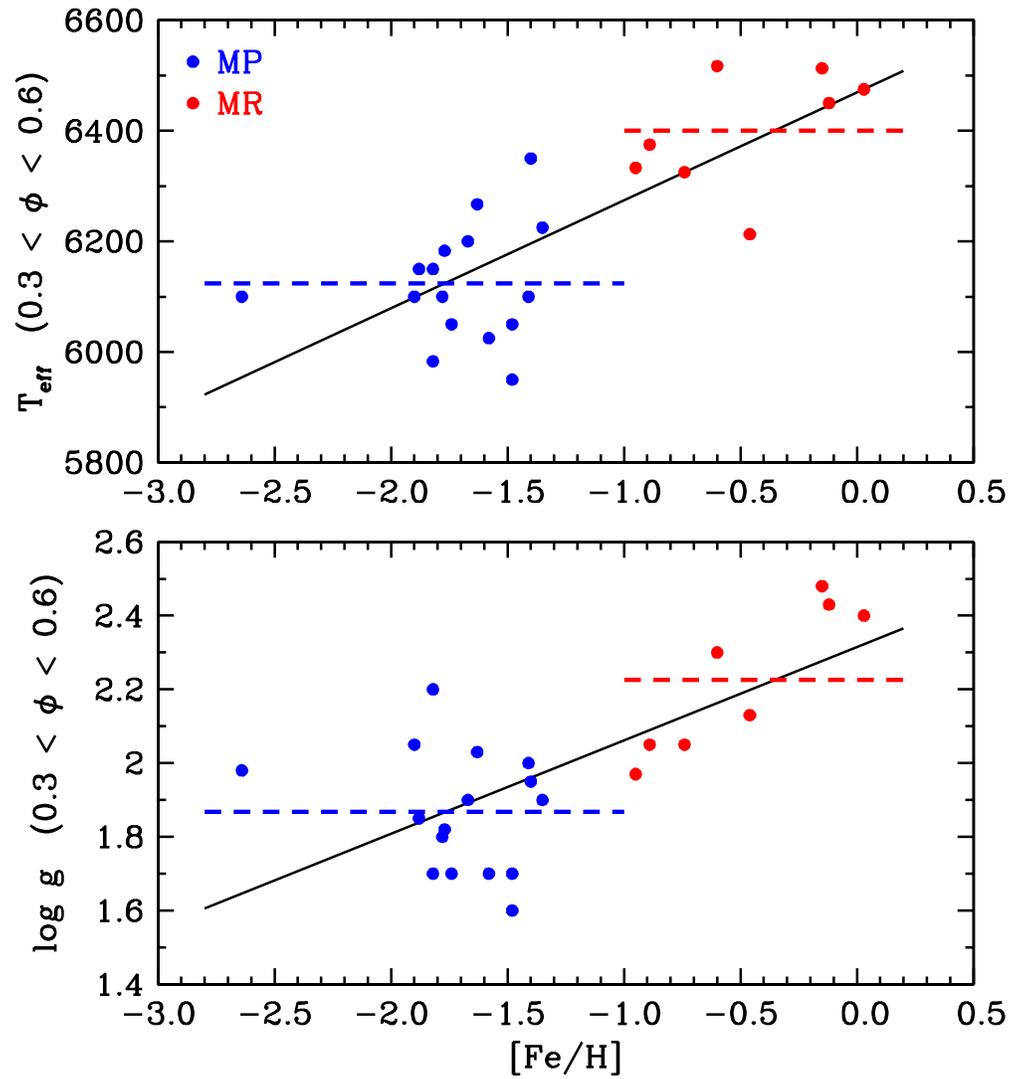}
\caption{\label{f4}
\footnotesize
Temperatures (upper panel) and gravities (lower panel) of program stars
plotted $versus$ their metallicities.
The symbols are as in Figure~\ref{f1}.
Dashed horizontal lines in each panel depict the mean \teff\ and \logg\
values of metal-rich and metal-poor stars.
The solid black lines represent linear regression fits to all the data 
points.}
\end{figure}

\clearpage
\begin{figure}
\epsscale{1.00}
\plotone{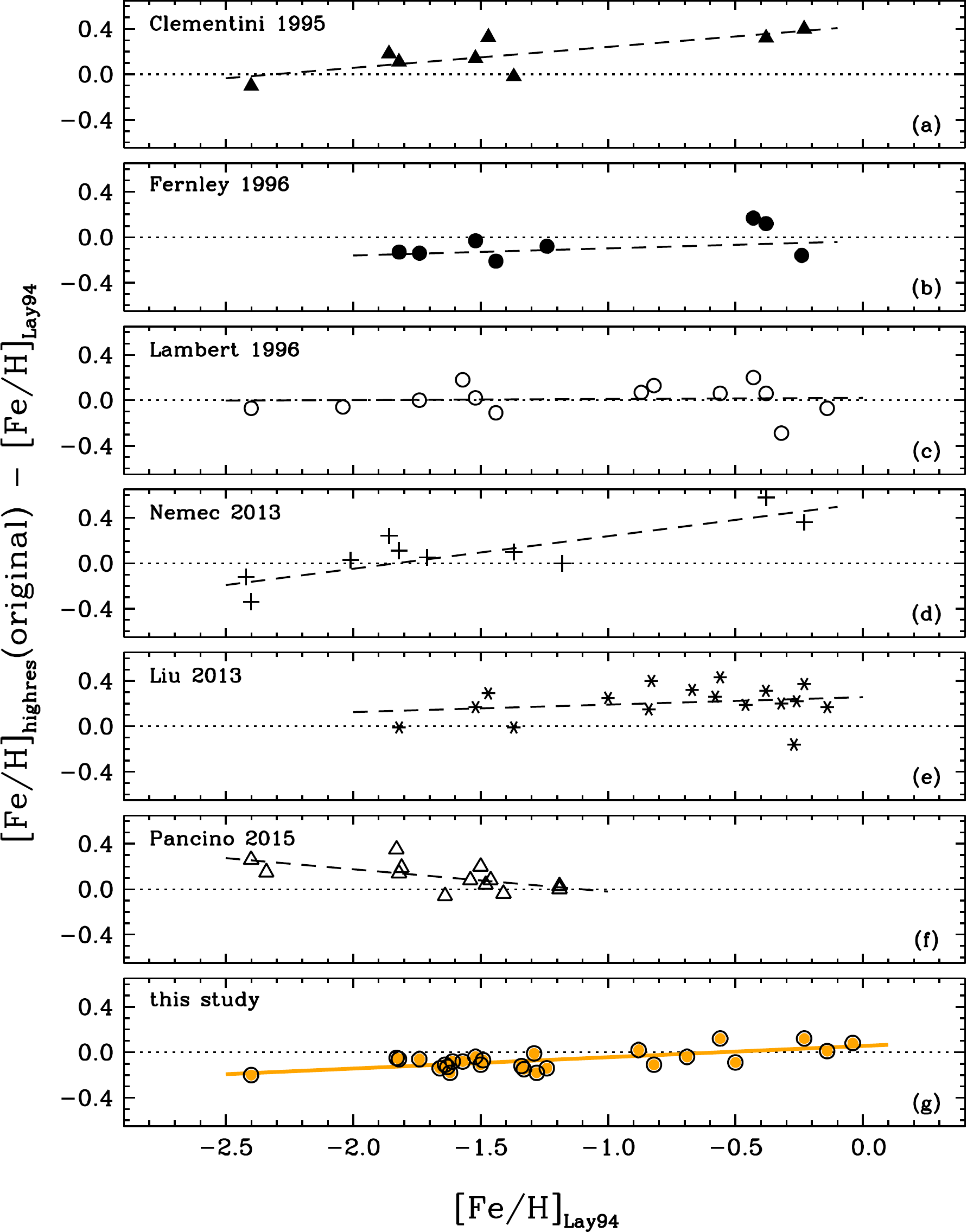}
\caption{
\label{f5} \footnotesize
Differences between RR~Lyr[Fe/H] ``original'' metallicities reported in 
seven high-resolution spectroscopic studies and those derived by Lay9 
from calibrated $\Delta S$ indices, plotted as functions of the Lay94 values.
The symbols used in this figure to distinguish among the different
studies will be also be used in Figures~\ref{f6} and \ref{f7}.
Sources briefly named in the figure panels are: (a), \cite{clementini95},
(b) \cite{fernley96}, (c) \cite{lambert96}, (d) \cite{nemec13},
(e) \cite{liu13}, (f) \cite{pancino15}, and (g) this study.
The orange line in the bottom panel represents a linear regression 
fit to the metallicities derived here.}
\end{figure}

\clearpage
\begin{figure}
\epsscale{1.00}
\plotone{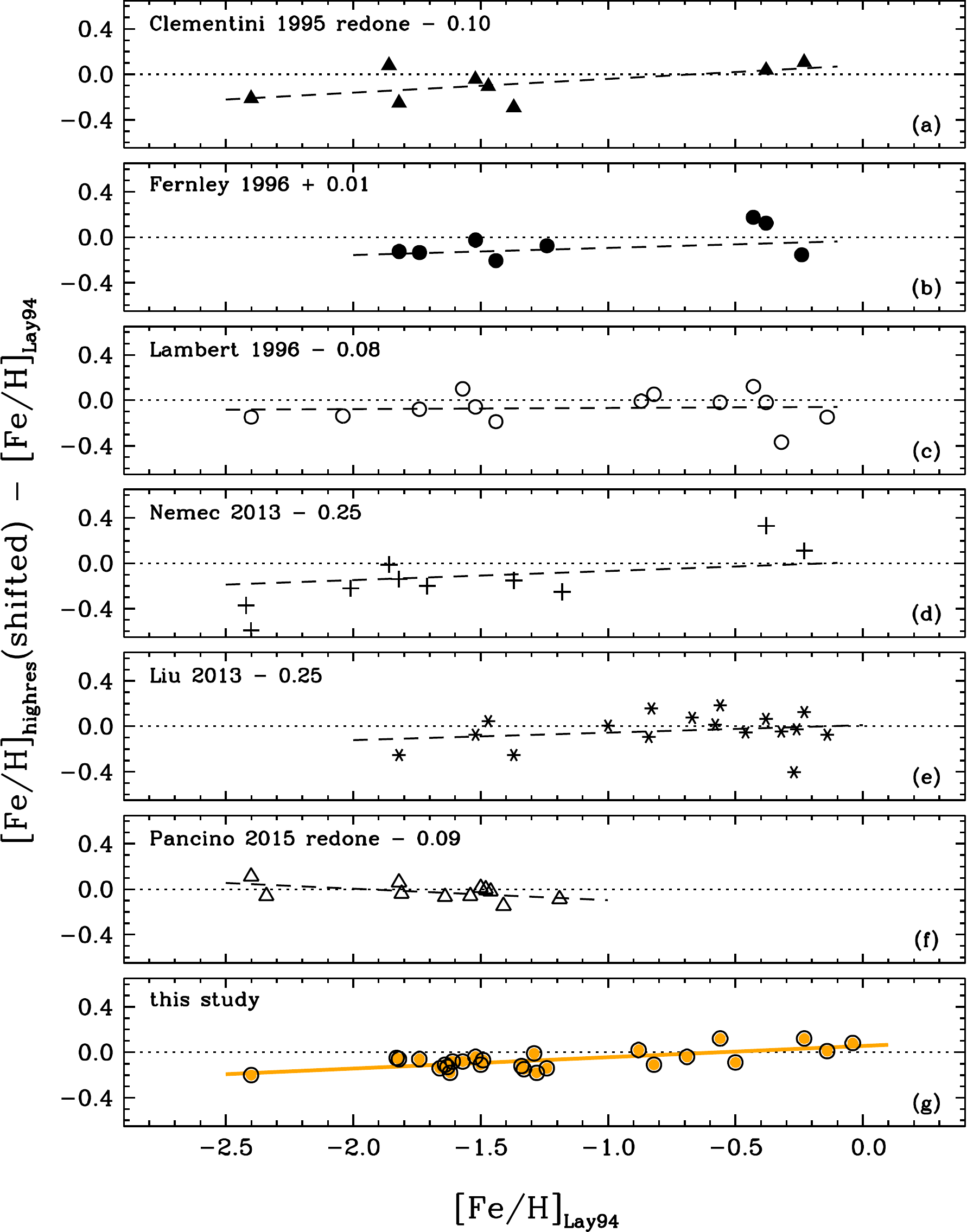}
\caption{
\label{f6} \footnotesize
Another comparision between high-resolution spectroscopic metallicities
and the Lay94 $\Delta S$ metallicities, but after reanalysis of
the $EW$s of \cite{clementini95} and \cite{pancino15}, and application of 
additive constants to bring agreement between the literature results and those
of this study at [Fe/H]~=~$-$1.25.
See the text for details of this procedure.
The symbols are as in Figure~\ref{f5}. }
\end{figure}

\clearpage
\begin{figure}
\epsscale{1.00}
\plotone{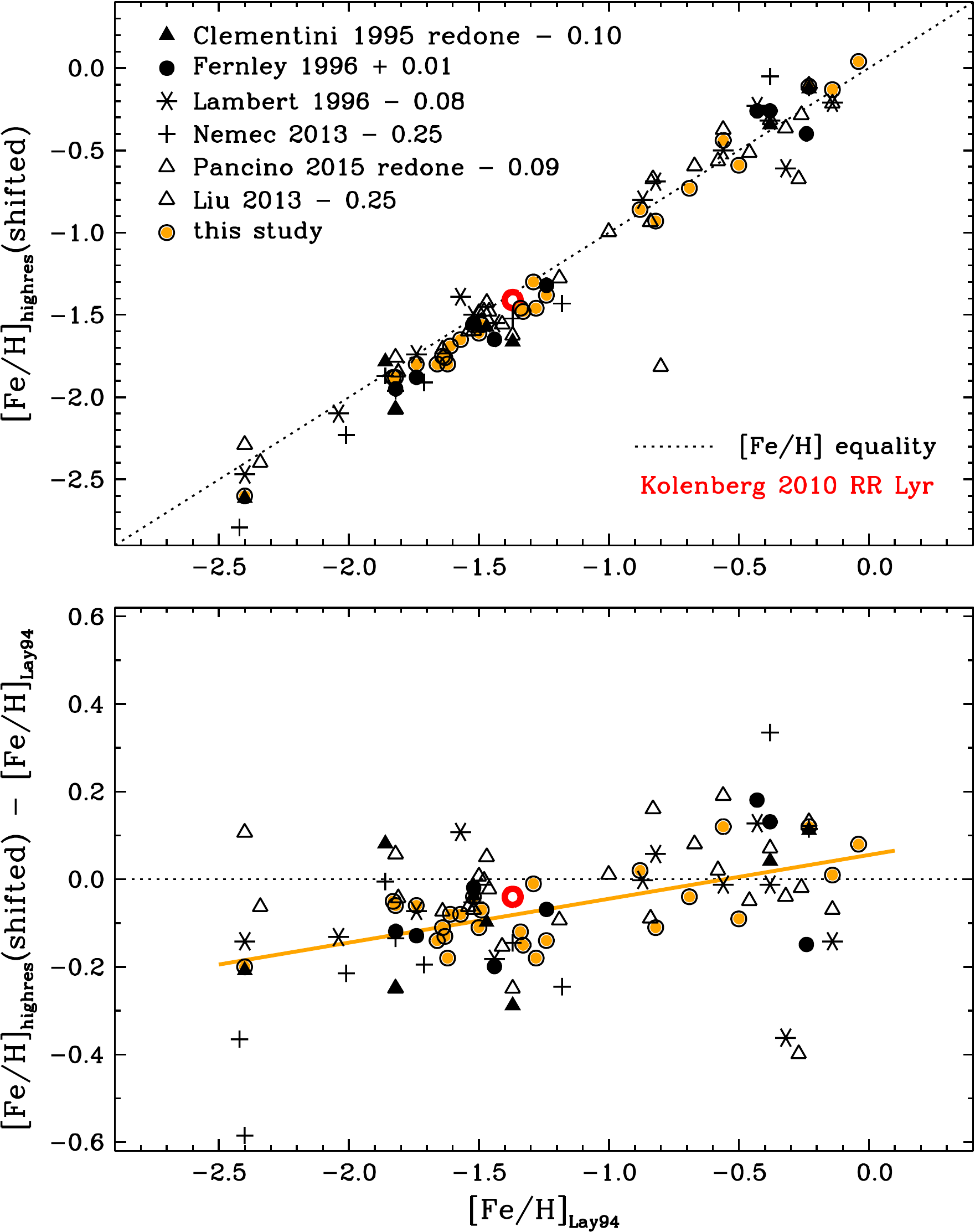}
\caption{
\label{f7} \footnotesize
Final comparison between high-resolution spectroscopic metallicities
and those of Lay94, after the reanalyses and scale shifts
as descirbed in the text have been applied.
Symbols (consistent with those of Figure~\ref{f5}) are redefined in 
the upper panel.
A red circle has been added to denote the metallicity for the
star RR~Lyr that was derived by \cite{kolenberg10}.
}
\end{figure}

\clearpage
\begin{figure}
\epsscale{1.00}
\plotone{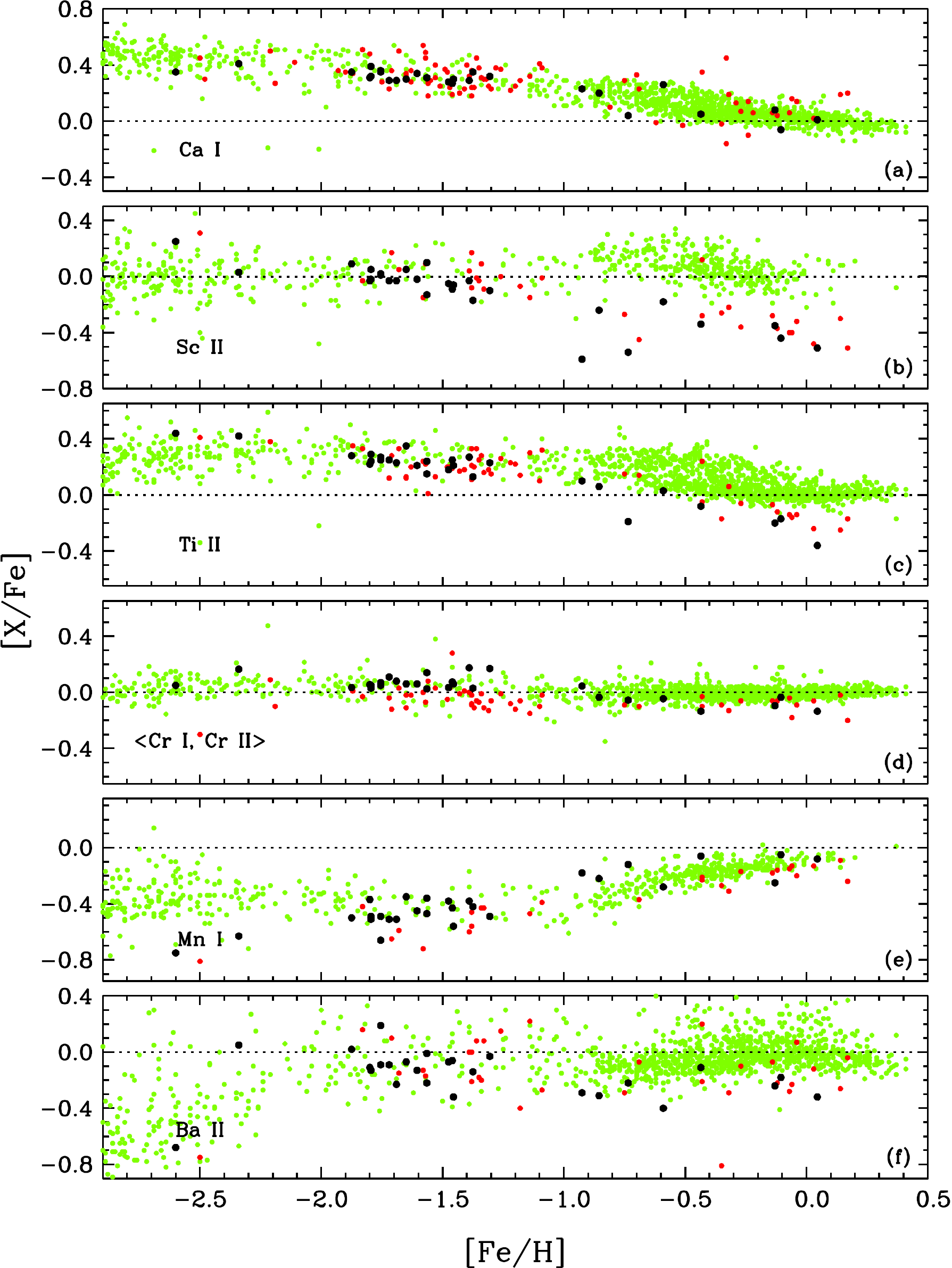}
\caption{
\label{f8} \footnotesize
Relative abundance ratios as functions of [Fe/H] metallicity from
our study (black dots), other RR~Lyr publications (red dots), and
from several very large-sample abundance surveys of non-variable
main sequence, subgiant, and red giant stars.
See text for sources of the external data.
}
\end{figure}

\clearpage
\figsetstart
\figsetnum{9}
\figsettitle{Velocities, Halpha Fluxes, and V Magnitudes for the Program Stats}

\figsetgrpstart
\figsetgrpnum{9.1}
\figsetgrptitle{Stars 1-4}
\figsetplot{f9_1.pdf}
\figsetgrpnote{Data for the Program Stars}
\figsetgrpend

\figsetgrpstart
\figsetgrpnum{9.2}
\figsetgrptitle{Stars 5-8}
\figsetplot{f9_2.pdf}
\figsetgrpnote{Data for the Program Stars}
\figsetgrpend

\figsetgrpstart
\figsetgrpnum{9.3}
\figsetgrptitle{Stars 9-12}
\figsetplot{f9_3.pdf}
\figsetgrpnote{Data for the Program Stars}
\figsetgrpend

\figsetgrpstart
\figsetgrpnum{9.4}
\figsetgrptitle{Stars 13-16}
\figsetplot{f9_4.pdf}
\figsetgrpnote{Data for the Program Stars}
\figsetgrpend

\figsetgrpstart
\figsetgrpnum{9.5}
\figsetgrptitle{Stars 17-20}
\figsetplot{f9_5.pdf}
\figsetgrpnote{Data for the Program Stars}
\figsetgrpend

\figsetgrpstart
\figsetgrpnum{9.6}
\figsetgrptitle{Stars 21-24}
\figsetplot{f9_6.pdf}
\figsetgrpnote{Data for the Program Stars}
\figsetgrpend

\figsetend
\begin{figure}                                                
\epsscale{1.10}                                               
\plotone{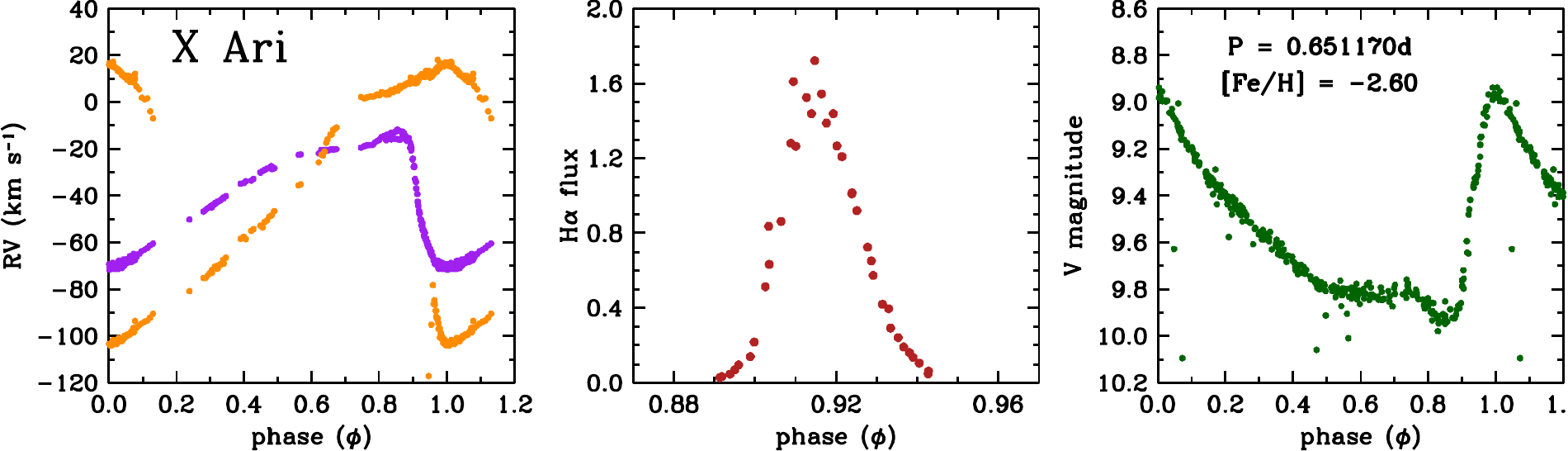}                                   
\caption{                                                     
\label{f9} \footnotesize                           
Variations in metallic-line (purple points) and H$\alpha$ (orange points)
radial velocities (left panel),   
H$\alpha$ emission (middle panel), and $V$-band magnitude as functions      
of pulsational phase.                                         
Only the data for X~Ari are shown here; the plots for the other
23 non-Blazhko program stars are available in the on-line version
of this paper.                                                
}                                                             
\end{figure}  
\clearpage
\begin{figure}
\epsscale{1.00}
\plotone{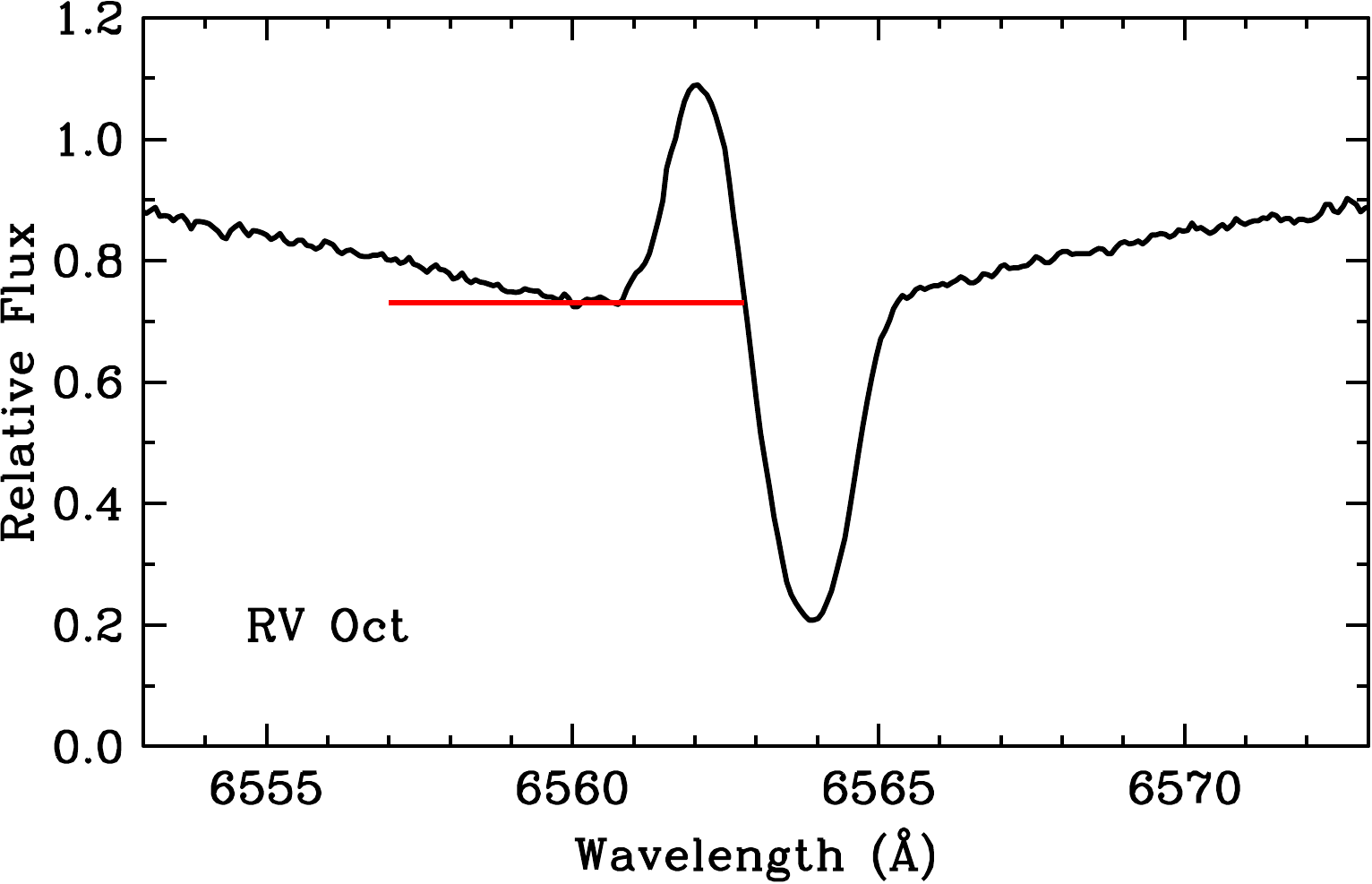}
\caption{\label{f10}
\footnotesize
H$\alpha$ emission flux at phase $\phi$~=~0.943 in RV Oct is measured 
as the sum of flux counts above the reference level indicated by the 
horizontal red line.}
\end{figure}

\clearpage
\begin{figure}
\epsscale{1.05}
\plotone{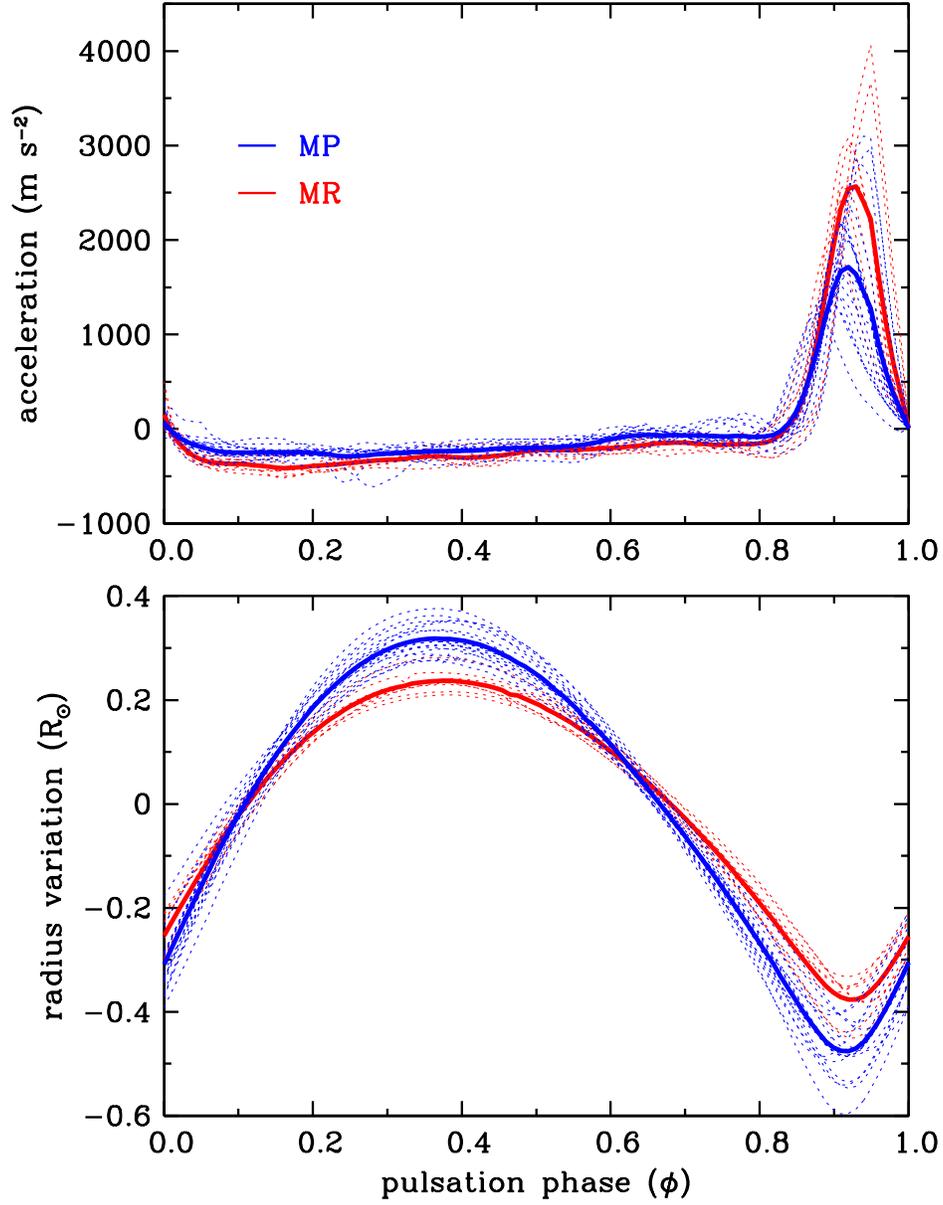} 
\caption{\label{f11}
\footnotesize
Dynamical acceleration and radius variation curves
for metal-rich RR\,ab stars (MR; red lines) and metal-poor RR\,ab stars 
(MP; blue lines).
The thin dashed lines show the accelerations and radius variations for
individual stars, and the thick solid lines are for simple means constructed
of the MR and MP data.}
\end{figure}

\clearpage
\begin{figure}
\epsscale{1.10}
\plotone{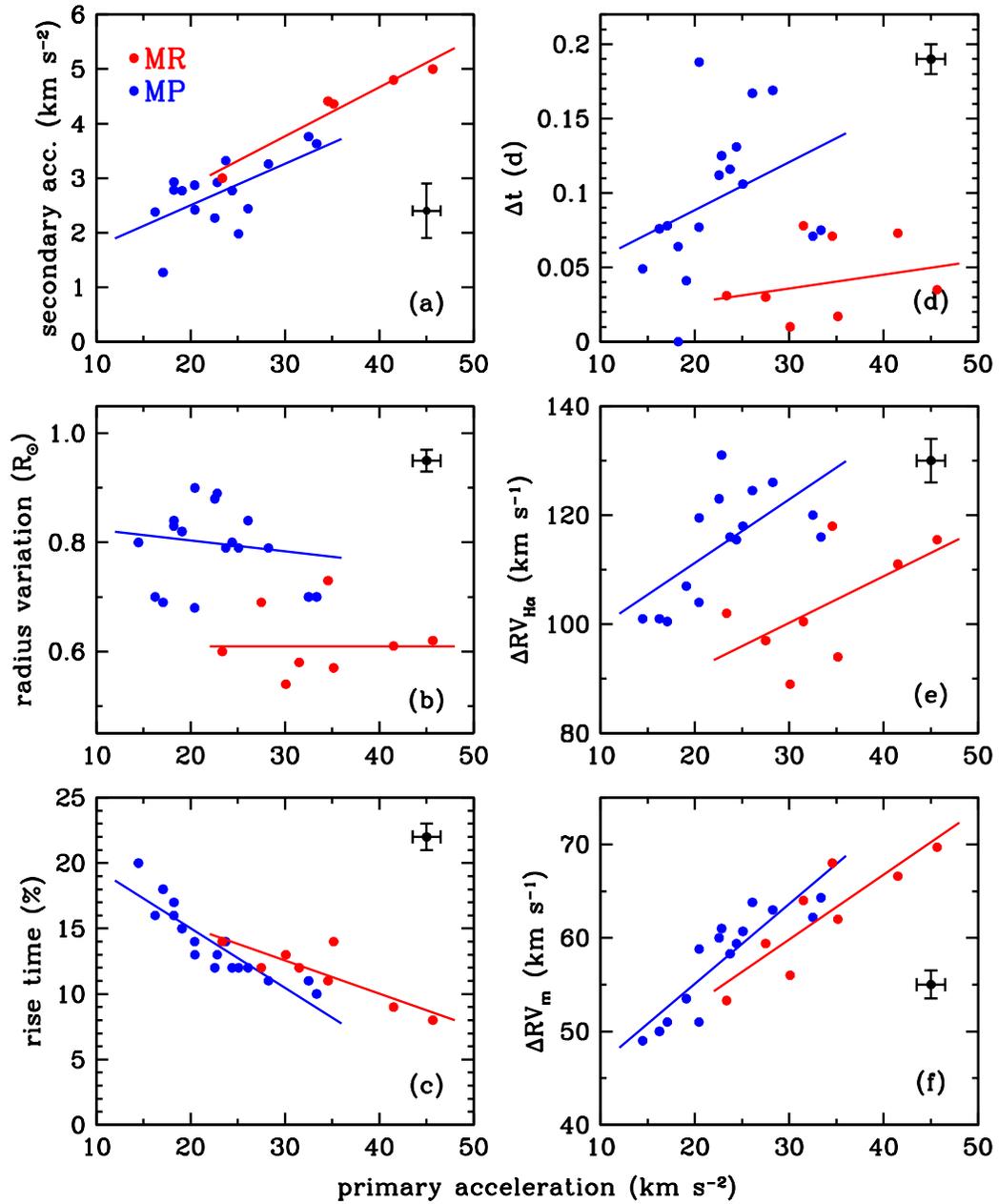}
\caption{\label{f12}
\footnotesize
Secondary acceleration, radius variation, rise time, duration of 
H$\alpha$ doubling, H$\alpha$ RV amplitude and metal RV amplitude 
$versus$ primary acceleration for metal-poor (blue) and metal-rich (red)
RRab stars.
Uncertainty estimates for the quantities are plotted int he figure 
captions; see Table~\ref{tab-rv}.}
\end{figure}

\clearpage
\begin{figure}
\epsscale{1.00}
\plotone{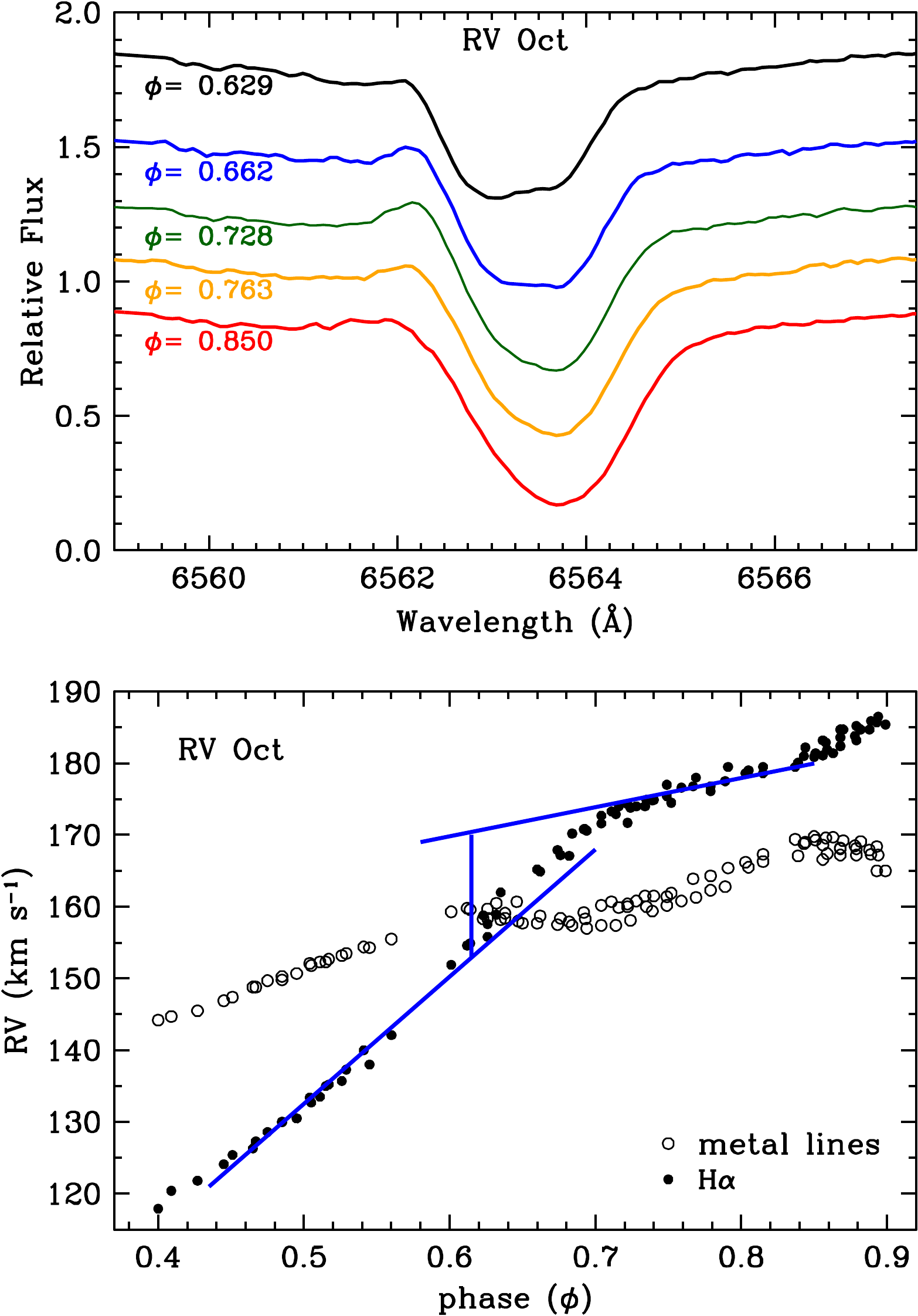}
\caption{\label{f13}
\footnotesize
The top panel contains a sequence of H$\alpha$ profiles during the $Sh_{PM3}$
 shock in the phase interval 0.63~$<$~$\phi$~$<$~0.85.  
The bottom panel shows metal (open black) and H$\alpha$ (solid black) radial velocities 
in this phase interval. 
The blue vertical line marks a 20 \kmsec\ shock discontinuity in H$\alpha$ 
radial velocity, evident in the line profiles at left.}
\end{figure}

\begin{figure}
\epsscale{1.10}
\plotone{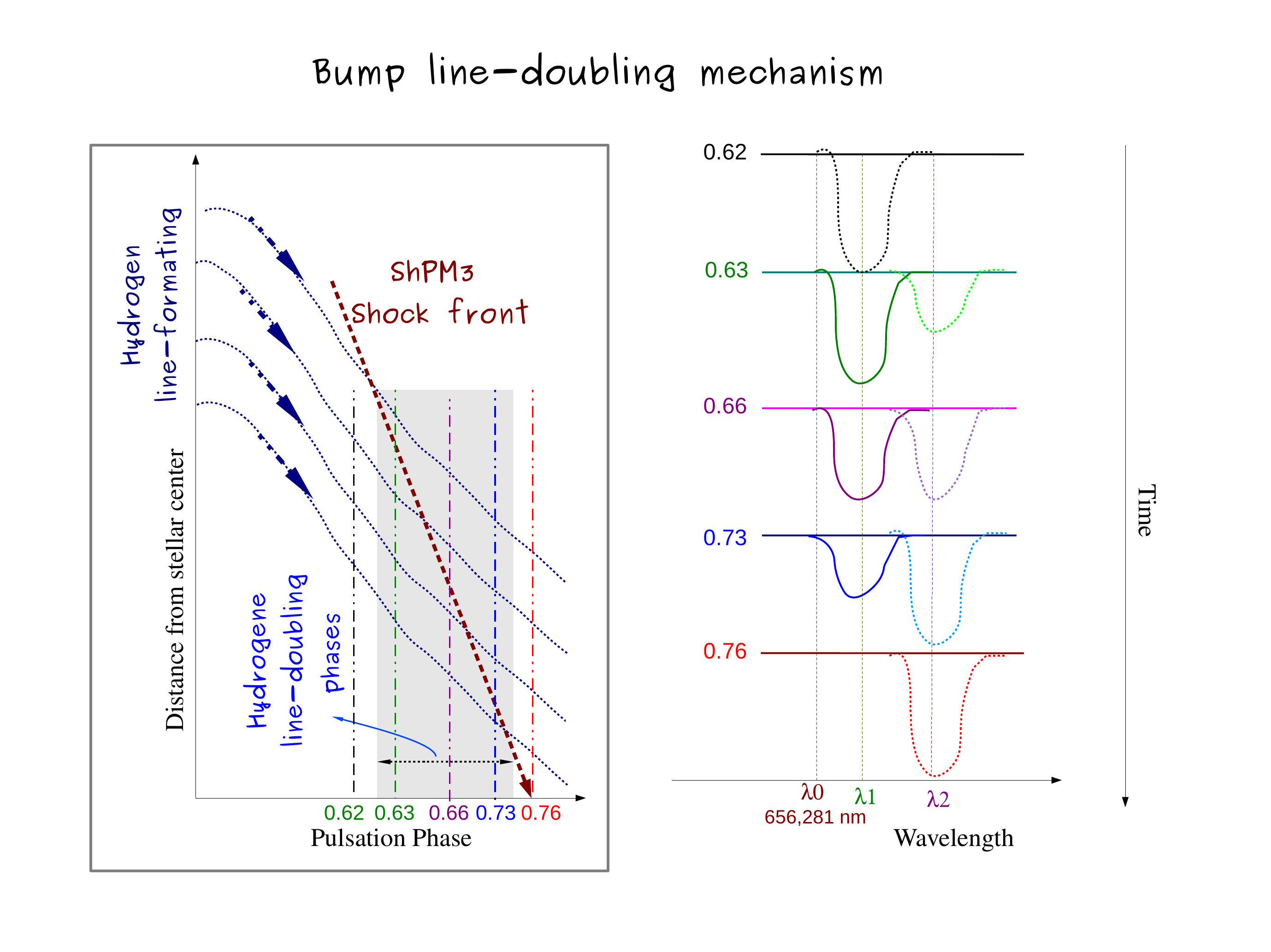} 
\caption{\label{f14}
\footnotesize
Bump line--doubling mechanism: temporal sequence followed by the intensity of both red components of  H$\alpha$ line--doubling during the bump in the light curve, close to the light minimum, 
when the $Sh_{PM3}$ ballistic shock wave propagates through the inward motion photosphere without the presence of any complicated
radiative process du to the shock wave. We assume in this line--doubling mechanism that the four layers have an inward motion with
a quasi--equal velocity towards the center of the star. An inward shock appears, with a greater velocity, 
progressively gets across the layers and  changes their velocity  to an higher velocity, inducing a small red--shifted component
that gradually increases when the shock deeply gets through the layers.}
\end{figure} 

\clearpage
\begin{figure}
\epsscale{1.10}
\plotone{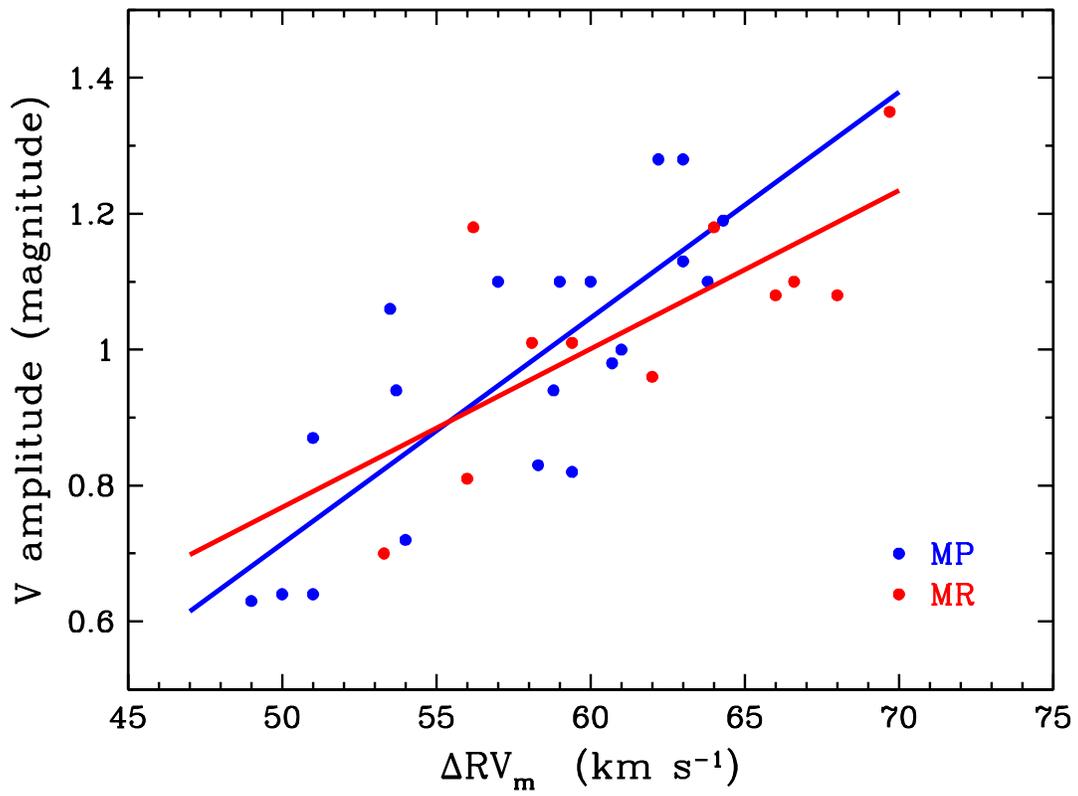} 
\caption{\label{f15}
\footnotesize
The correlations between $V$ light amplitude and metallic line (photospheric) 
radial velocity amplitude for metal-poor (blue) and metal-rich (red) 
RRab stars.}
\end{figure}

\clearpage
\begin{figure}
\epsscale{1.00}
\plotone{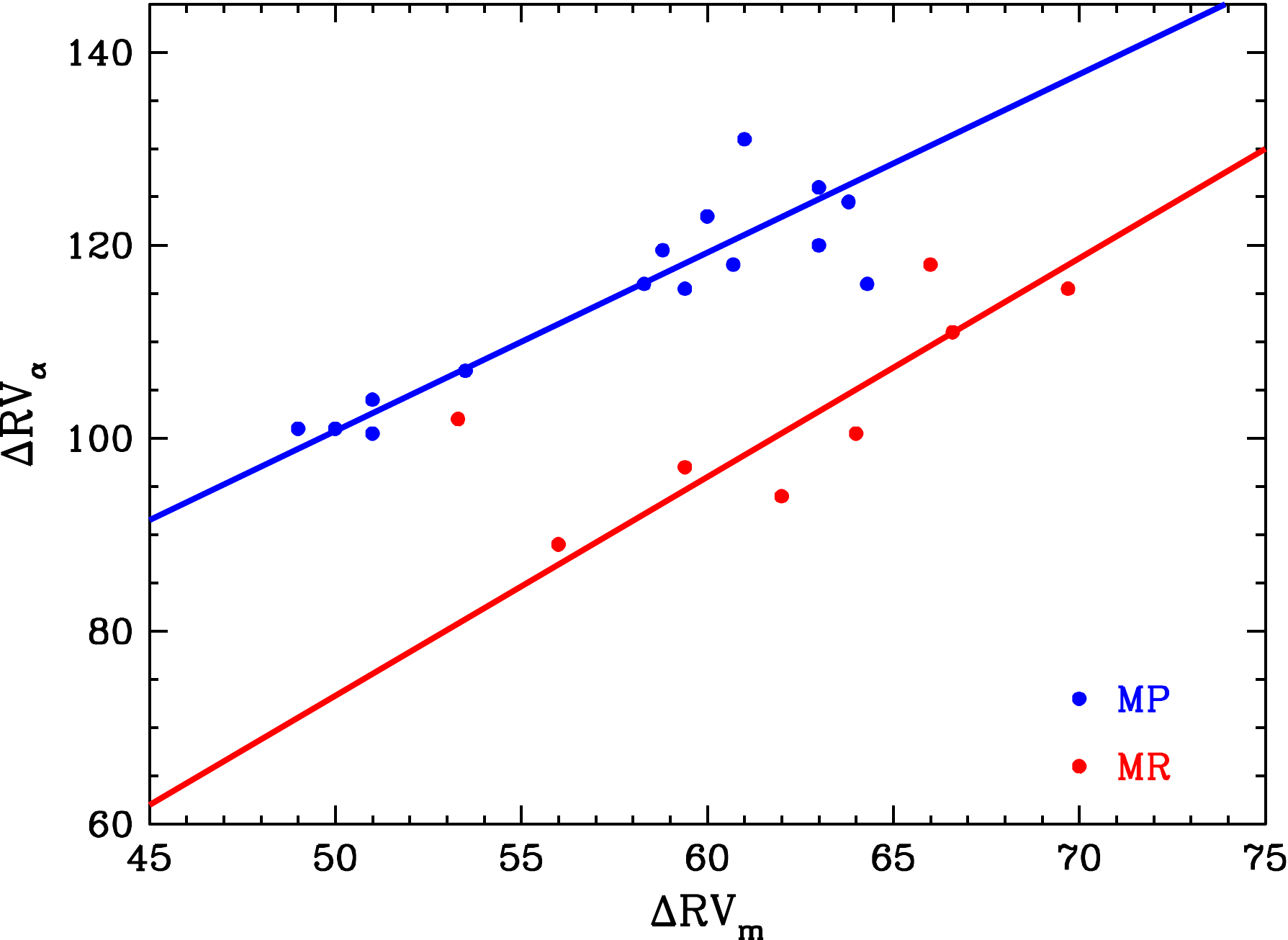} 
\caption{\label{f16}
\footnotesize
A plot of H$\alpha$ RV amplitude $versus$ metal RV amplitude for 
metal-poor (blue) and metal-rich (red) RRab stars.}
\end{figure}  

\clearpage
\begin{figure}
\epsscale{1.00}
\plotone{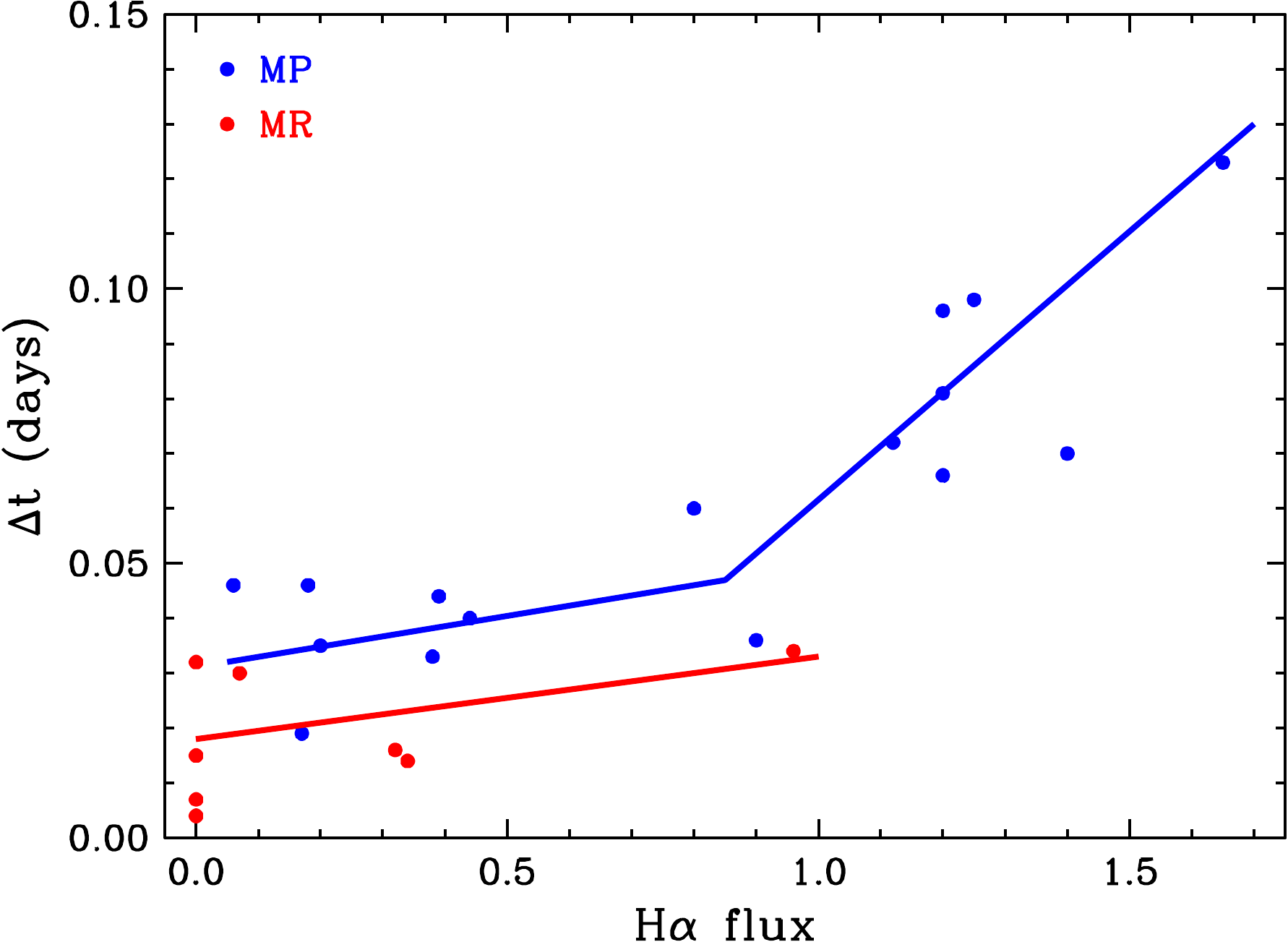}
\caption{\label{f17}
\footnotesize
Duration of H$\alpha$ doubling $versus$ H$\alpha$ peak flux for metal-poor 
(blue) and metal--rich (red) RRab stars.
H$\alpha$ flux is measured in units of the normalized continuum flux at 
6500~\AA.}
\end{figure}

\clearpage
\begin{figure}
\epsscale{1.00}
\plotone{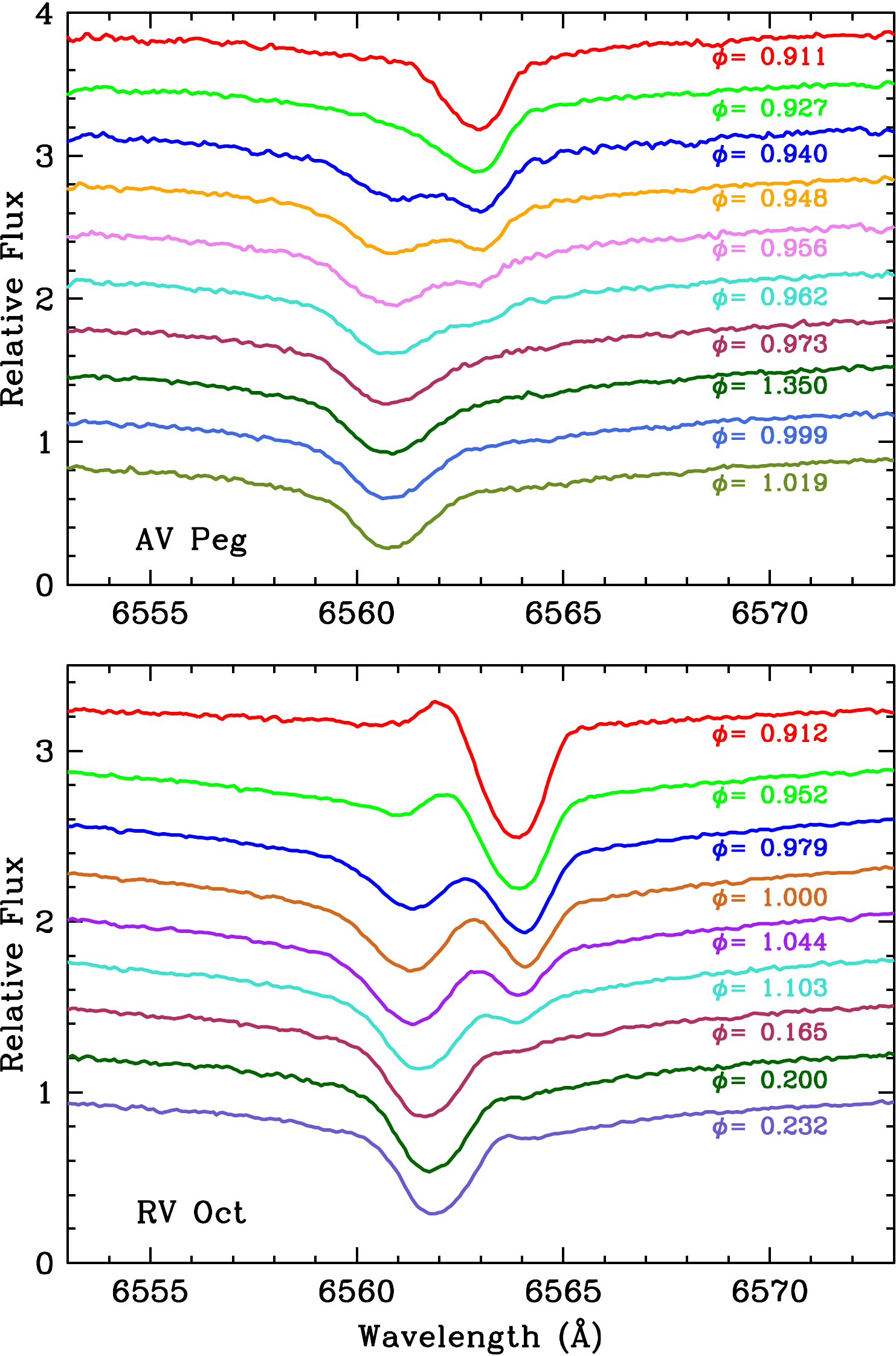}
\caption{\label{f18}
\footnotesize
Spectra of the H$\alpha$ line profile as selected phases of metal-rich 
AV~Peg (top panel) and metal-poor RV Oct (bottom panel).}
\end{figure}

\begin{figure}
\epsscale{1.00}
\plotone{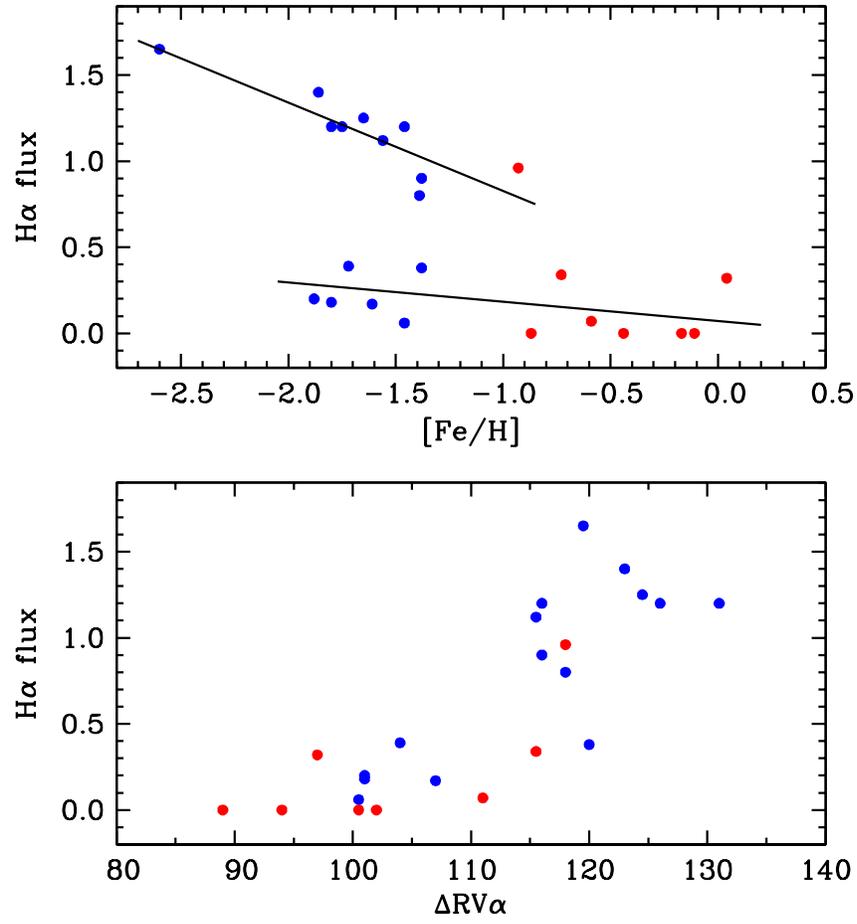}
\caption{\label{f19}
\footnotesize
$H\alpha$ emission as function of metallicity (top panel) and hydrogen 
radial velocity amplitude (bottom panel).
H$\alpha$ flux is measured in units of the normalized continuum flux at 
6500~\AA.}
\end{figure}


  \clearpage
\begin{center}
\begin{deluxetable}{lcrrrrrrr}
\tabletypesize{\footnotesize}
\tablewidth{0pt}
\tablecaption{Program Star Data\label{tab-stars}}
\tablecolumns{9}
\tablehead{
\colhead{Name}                     &
\colhead{Name}                     &
\colhead{RA$_{2000}$}              &
\colhead{Dec$_{2000}$}             &
\colhead{$P$}                      &
\colhead{$T_0$}                    &
\colhead{$V$}                      &
\colhead{$V_{amp}$}                &
\colhead{[Fe/H]}                   \\
\colhead{GCVS}                     &
\colhead{ASAS\tablenotemark{a}}    &
\colhead{ASAS}                     &
\colhead{ASAS}                     &
\colhead{ASAS}                     &
\colhead{ASAS}                     &
\colhead{}                         &
\colhead{}                         &
\colhead{Lay94\tablenotemark{b}}
}
\startdata
WY Ant    & 101605-2943.7 &  10:16:05 & $-$29:43:42 & 0.574340 & 1870.76 & 10.37 & 0.85 & $-$1.66 \\
BS Aps    & 162051-7140.3 &  16:20:51 & $-$71:40:18 & 0.582577 & 1915.69 & 11.90 & 0.68 & $-$1.33 \\
XZ Aps    & 145205-7940.7 &  14:52:05 & $-$79:40:42 & 0.587300 & 1887.15 & 11.94 & 1.10 & $-$1.57 \\
DN Aqr    & 231917-2413.0 &  23:19:17 & $-$24:13:00 & 0.633758 & 1872.33 & 10.85 & 0.72 & $-$1.63 \\
SW Aqr    & 211518+0004.6 &  21:15:18 &     0:04:36 & 0.459300 & 1876.17 & 10.56 & 1.28 & $-$1.24 \\
X  Ari    & 030831+1026.8 &   3:08:31 &    10:26:48 & 0.651172 & 1890.10 &  9.24 & 0.94 & $-$2.40 \\
RR Cet    & 013208+0120.5 &   1:32:08 &     1:20:30 & 0.553030 & 2143.63 &  9.26 & 0.82 & $-$1.52 \\
W  Crt    & 112630-1754.9 &  11:26:30 & $-$17:54:54 & 0.412011 & 1871.64 & 10.90 & 1.10 & $-$0.50 \\
DX Del    & 204729+1227.8 &  20:47:29 &    12:27:48 & 0.472650 & 2415.85 &  9.81 & 0.70 & $-$0.56 \\
SX For    & 033022-3603.2 &   3:30:22 & $-$36:03:12 & 0.605350 & 1870.42 & 10.89 & 0.64 & $-$1.62 \\
DT Hya    & 115400-3115.7 &  11:54:00 & $-$31:15:42 & 0.567970 & 1872.11 & 12.53 & 0.98 & \nodata \\
V  Ind    & 211130-4504.5 &  21:11:30 & $-$45:04:30 & 0.479600 & 1873.49 &  9.62 & 1.06 & $-$1.50 \\
SS Leo    & 113355-0002.0 &  11:33:55 & $-$00:02:00 & 0.626330 & 1873.06 & 10.47 & 1.00 & $-$1.83 \\
ST Leo    & 113833+1033.7 &  11:38:33 &    10:33:42 & 0.477983 & 2384.94 & 10.99 & 1.19 & $-$1.29 \\
Z  Mic    & 211623-3017.0 &  21:16:23 & $-$30:17:00 & 0.586930 & 1873.36 & 11.32 & 0.64 & $-$1.28 \\
RV Oct    & 134632-8424.1 &  13:46:32 & $-$84:24:06 & 0.571184 & 1891.66 & 10.53 & 1.13 & $-$1.34 \\
UV Oct    & 163225-8354.2 &  16:32:25 & $-$83:54:12 & 0.542561 & 1890.60 &  9.19 & 0.82 & $-$1.61 \\
V0445 Oph & 162441-0632.5 &  16:24:41 & $-$06:32:30 & 0.397030 & 1939.03 & 10.54 & 0.81 & $-$0.23 \\
AV Peg    & 215203+2234.4 &  21:52:03 &    22:34:24 & 0.390385 & 2758.11 &  9.95 & 0.96 & $-$0.14 \\
HH Pup    & 072036-4642.5 &   7:20:36 & $-$46:42:30 & 0.390745 & 1869.65 & 10.57 & 1.24 & $-$0.69 \\
AN Ser    & 155331+1257.6 &  15:53:31 &    12:57:36 & 0.522080 & 2701.17 & 10.46 & 1.01 & $-$0.04 \\
VY Ser    & 153102+0141.0 &  15:31:02 &     1:41:00 & 0.714100 & 1922.71 &  9.84 & 0.63 & $-$1.82 \\
v1645 Sgr & 202044-4107.1 &  20:20:44 & $-$41:07:06 & 0.552979 & 1876.53 & 10.99 & 0.84 & $-$1.74 \\
W  Tuc    & 005810-6323.8 &   0:58:10 & $-$63:23:48 & 0.642260 & 1869.52 & 10.90 & 1.11 & $-$1.64 \\
CD Vel    & 094438-4552.6 &   9:44:38 & $-$45:52:36 & 0.573510 & 1869.91 & 11.66 & 0.87 & \nodata \\
AS Vir    & 125246-1015.6 &  12:52:46 & $-$10:15:36 & 0.553439 & 1886.61 & 11.66 & 0.72 & $-$1.49 \\
ST Vir    & 142739-0054.1 &  14:27:39 & $-$00:54:06 & 0.410806 & 1906.92 & 11.00 & 1.18 & $-$0.88 \\
UU Vir    & 120835-0027.4 &  12:08:35 & $-$00:27:24 & 0.475597 & 1886.54 & 10.06 & 1.08 & $-$0.82 \\
\enddata
\tablenotetext{a}{All Skay Automated Survey \citep{pojmanski03}}
\tablenotetext{b}{Lay94}
\end{deluxetable}
\end{center}

\clearpage
\begin{center}
\begin{deluxetable}{lccc}
\tabletypesize{\footnotesize}
\tablewidth{0pt}
\tablecaption{Standard deviations of the regressions\label{tab-sigmas}}
\tablecolumns{4}
\tablehead{
Quantity          &
MP                & 
MR                & 
ALL
}
\startdata
\teff\      &   102  &   100  &   108  \\
\logg\      &  0.16  &  0.19  &  0.18  \\
\enddata
\end{deluxetable}
\end{center}

\clearpage
\begin{center}
\begin{deluxetable}{lcrrrrrrrrrrr}
\tabletypesize{\footnotesize}
\tablewidth{0pt}
\tablecaption{Model Atmosphere Parameters\label{tab-models}}
\tablecolumns{13}
\tablehead{
\colhead{Star}                     &
\colhead{[Fe/H]}                   &
\colhead{$\phi$}                   &
\colhead{\teff}                    &
\colhead{\logg}                    &
\colhead{\vmicro}                  &
\colhead{[M/H]}                    &
\colhead{[Fe/H]}                   &
\colhead{$\sigma$}                 &
\colhead{\#}                       &
\colhead{[Fe/H]}                   &
\colhead{$\sigma$}                 &
\colhead{\#}                       \\
\colhead{}                         &
\colhead{Lay94}                    &
\colhead{}                         &
\colhead{K}                        &
\colhead{}                         &
\colhead{\kmsec}                   &
\colhead{model}                    &
\colhead{{\sc i}}                  &
\colhead{{\sc i}}                  &
\colhead{{\sc i}}                  &
\colhead{{\sc ii}}                 &
\colhead{{\sc ii}}                 &
\colhead{{\sc ii}}                  
}
\startdata
    WY Ant  & -1.66 & 0.50 & 6150 & 2.20 & 3.0 & -1.80 & -1.82 & 0.14 & 106 & -1.77 & 0.14 &  24 \\
    BS Aps  & -1.33 & 0.50 & 6000 & 1.80 & 3.0 & -1.50 & -1.51 & 0.11 & 106 & -1.44 & 0.14 &  20 \\
    XZ Aps  & -1.57 & 0.50 & 6200 & 1.90 & 2.8 & -1.60 & -1.65 & 0.15 & 110 & -1.65 & 0.14 &  23 \\
    DN Aqr  & -1.63 & 0.25 & 6100 & 1.80 & 3.0 & -1.70 & -1.80 & 0.15 & 104 & -1.75 & 0.17 &  28 \\
    DN Aqr  & -1.63 & 0.37 & 6100 & 1.80 & 2.8 & -1.70 & -1.74 & 0.13 & 129 & -1.73 & 0.12 &  33 \\
    SW Aqr  & -1.29 & 0.28 & 6500 & 1.90 & 2.9 & -1.40 & -1.41 & 0.15 & 127 & -1.40 & 0.11 &  32 \\
    SW Aqr  & -1.29 & 0.41 & 6200 & 2.00 & 2.9 & -1.40 & -1.37 & 0.20 &  90 & -1.32 & 0.18 &  17 \\
     X Ari  & -2.40 & 0.19 & 6650 & 2.30 & 2.7 & -2.50 & -2.54 & 0.10 &  35 & -2.53 & 0.21 &  14 \\
     X Ari  & -2.40 & 0.30 & 6200 & 1.90 & 2.8 & -2.60 & -2.65 & 0.11 &  56 & -2.66 & 0.26 &  20 \\
     X Ari  & -2.40 & 0.37 & 6100 & 2.15 & 2.8 & -2.60 & -2.63 & 0.10 &  63 & -2.59 & 0.11 &  20 \\
     X Ari  & -2.40 & 0.47 & 6000 & 1.90 & 2.8 & -2.60 & -2.57 & 0.11 &  72 & -2.60 & 0.14 &  20 \\
    RR Cet  & -1.52 & 0.34 & 6100 & 1.70 & 2.9 & -1.50 & -1.50 & 0.13 & 124 & -1.48 & 0.13 &  30 \\
    RR Cet  & -1.52 & 0.55 & 5950 & 1.70 & 3.1 & -1.50 & -1.63 & 0.14 & 116 & -1.63 & 0.14 &  32 \\
     W Crt  & -0.50 & 0.22 & 6850 & 2.00 & 2.9 & -0.50 & -0.65 & 0.14 & 113 & -0.61 & 0.20 &  20 \\
     W Crt  & -0.50 & 0.32 & 6650 & 2.10 & 2.8 & -0.60 & -0.58 & 0.13 & 111 & -0.61 & 0.14 &  21 \\
     W Crt  & -0.50 & 0.40 & 6400 & 2.20 & 3.0 & -0.60 & -0.61 & 0.12 &  99 & -0.59 & 0.16 &  19 \\
     W Crt  & -0.50 & 0.58 & 6500 & 2.60 & 2.5 & -0.50 & -0.55 & 0.21 & 101 & -0.52 & 0.24 &  23 \\
    DX Del  & -0.56 &-0.25 & 6300 & 2.00 & 3.2 & -0.50 & -0.44 & 0.14 &  84 & -0.47 & 0.20 &  15 \\
    DX Del  & -0.56 & 0.35 & 6150 & 2.00 & 3.1 & -0.60 & -0.51 & 0.15 &  83 & -0.56 & 0.16 &  16 \\
    DX Del  & -0.56 & 0.42 & 6150 & 2.40 & 3.2 & -0.50 & -0.31 & 0.19 &  68 & -0.36 & 0.28 &  10 \\
    DX Del  & -0.56 & 0.57 & 6250 & 2.10 & 3.4 & -0.50 & -0.39 & 0.12 &  67 & -0.46 & 0.40 &  13 \\
    SX For  & -1.62 & 0.31 & 6000 & 1.70 & 2.7 & -1.80 & -1.79 & 0.11 & 128 & -1.79 & 0.15 &  33 \\
    SX For  & -1.62 & 0.36 & 6000 & 1.70 & 2.8 & -1.80 & -1.80 & 0.11 & 127 & -1.79 & 0.15 &  33 \\
    Sx For  & -1.62 & 0.45 & 5950 & 1.70 & 2.8 & -1.80 & -1.81 & 0.12 & 120 & -1.80 & 0.15 &  29 \\
    DT Hya  &\nodata& 0.50 & 6100 & 2.00 & 3.1 & -1.50 & -1.43 & 0.13 &  82 & -1.35 & 0.13 &  18 \\
     V Ind  & -1.50 & 0.32 & 6400 & 2.00 & 2.7 & -1.50 & -1.56 & 0.10 & 124 & -1.53 & 0.12 &  33 \\
     V Ind  & -1.50 & 0.40 & 6200 & 2.00 & 2.8 & -1.50 & -1.67 & 0.11 & 128 & -1.62 & 0.19 &  33 \\
     V Ind  & -1.50 & 0.47 & 6200 & 2.10 & 2.7 & -1.50 & -1.63 & 0.11 & 130 & -1.62 & 0.15 &  32 \\
     SSLeo  & -1.83 & 0.31 & 6200 & 2.10 & 2.9 & -1.90 & -1.87 & 0.12 & 113 & -1.86 & 0.15 &  28 \\
     SSLeo  & -1.83 & 0.41 & 6100 & 2.10 & 2.8 & -1.80 & -1.89 & 0.14 & 114 & -1.86 & 0.19 &  27 \\
     SSLeo  & -1.83 & 0.54 & 6100 & 2.10 & 3.0 & -1.80 & -1.88 & 0.17 &  87 & -1.82 & 0.13 &  22 \\
     SSLeo  & -1.83 & 0.56 & 6000 & 1.90 & 2.9 & -1.90 & -1.93 & 0.20 &  92 & -1.89 & 0.14 &  22 \\
   ST Leo   & -1.29 & 0.22 & 6650 & 2.00 & 3.0 & -1.30 & -1.25 & 0.21 & 106 & -1.28 & 0.19 &  28 \\
    ST Leo  & -1.29 & 0.32 & 6300 & 1.70 & 2.7 & -1.30 & -1.29 & 0.15 & 122 & -1.26 & 0.17 &  30 \\
    ST Leo  & -1.29 & 0.45 & 6150 & 2.10 & 2.8 & -1.50 & -1.41 & 0.17 & 130 & -1.34 & 0.24 &  30 \\
     Z Mic  & -1.28 & 0.50 & 5950 & 1.60 & 3.0 & -1.40 & -1.46 & 0.12 & 103 & -1.46 & 0.15 &  21 \\
    RV Oct  & -1.34 & 0.50 & 6050 & 1.70 & 2.8 & -1.50 & -1.47 & 0.12 & 112 & -1.44 & 0.17 &  22 \\
    UV Oct  & -1.61 & 0.50 & 6050 & 1.70 & 2.8 & -1.70 & -1.69 & 0.11 & 115 & -1.69 & 0.12 &  25 \\
  v445 Oph  & -0.23 & 0.28 & 6600 & 2.30 & 2.8 & -0.10 &  0.00 & 0.17 &  74 & -0.12 & 0.12 &  10 \\
  v445 Oph  & -0.23 & 0.34 & 6450 & 2.40 & 3.1 & -0.10 & -0.10 & 0.17 &  68 & -0.21 & 0.13 &  11 \\
  v445 Oph  & -0.23 & 0.44 & 6400 & 2.40 & 2.9 & -0.10 & -0.05 & 0.12 &  73 & -0.14 & 0.15 &  10 \\
  v445 Oph  & -0.23 & 0.53 & 6350 & 2.60 & 3.0 & -0.10 & -0.07 & 0.16 &  65 & -0.15 & 0.22 &   9 \\
    AV Peg  & -0.14 & 0.30 & 6600 & 2.30 & 2.8 & -0.10 & -0.13 & 0.14 &  95 & -0.17 & 0.17 &  17 \\
    AV Peg  & -0.14 & 0.36 & 6500 & 2.40 & 2.7 & -0.20 & -0.15 & 0.14 &  88 & -0.17 & 0.16 &  16 \\
    AV Peg  & -0.14 & 0.51 & 6400 & 2.60 & 2.8 & -0.20 & -0.18 & 0.19 &  70 & -0.19 & 0.15 &  14 \\
    AV Peg  & -0.14 & 0.57 & 6550 & 2.60 & 3.1 & -0.20 &  0.02 & 0.14 &  67 & -0.08 & 0.19 &  10 \\
   HH  Pup  & -0.69 & 0.13 & 7050 & 1.90 & 3.0 & -0.80 & -0.80 & 0.15 &  96 & -0.72 & 0.15 &  21 \\
    HH Pup  & -0.69 & 0.22 & 6750 & 2.00 & 3.0 & -0.80 & -0.73 & 0.10 & 104 & -0.71 & 0.15 &  19 \\
    HH Pup  & -0.69 & 0.35 & 6400 & 2.00 & 2.7 & -0.80 & -0.72 & 0.12 & 114 & -0.72 & 0.16 &  23 \\
    HH Pup  & -0.69 & 0.45 & 6250 & 2.10 & 2.7 & -0.80 & -0.73 & 0.14 & 112 & -0.73 & 0.17 &  23 \\
    AN Ser  & -0.04 & 0.22 & 6850 & 2.10 & 3.1 &  0.00 &  0.05 & 0.11 &  77 &  0.01 & 0.14 &  10 \\
    AN Ser  & -0.04 & 0.32 & 6550 & 2.20 & 3.0 & -0.10 &  0.08 & 0.13 &  70 &  0.03 & 0.19 &  10 \\
    AN Ser  & -0.04 & 0.58 & 6400 & 2.60 & 3.1 &  0.00 &  0.06 & 0.15 &  57 &  0.04 & 0.20 &  12 \\
    VY Ser  & -1.29 & 0.23 & 6200 & 1.85 & 2.9 & -1.80 & -1.92 & 0.09 & 105 & -1.90 & 0.14 &  28 \\
    VY Ser  & -1.29 & 0.29 & 6200 & 1.85 & 2.9 & -1.80 & -1.86 & 0.10 & 114 & -1.86 & 0.11 &  31 \\
    VY Ser  & -1.29 & 0.37 & 6100 & 1.85 & 2.8 & -1.70 & -1.86 & 0.10 & 123 & -1.86 & 0.13 &  31 \\
 1645v Sgr  & -1.74 & 0.50 & 6200 & 2.00 & 3.0 & -1.80 & -1.81 & 0.13 & 109 & -1.79 & 0.14 &  23 \\
     W Tuc  & -1.64 & 0.28 & 6350 & 1.75 & 3.0 & -1.70 & -1.75 & 0.14 &  91 & -1.73 & 0.11 &  27 \\
     W Tuc  & -1.64 & 0.40 & 6100 & 1.85 & 3.0 & -1.70 & -1.77 & 0.11 & 114 & -1.67 & 0.17 &  29 \\
     W Tuc  & -1.64 & 0.48 & 6100 & 1.85 & 3.0 & -1.70 & -1.79 & 0.13 & 108 & -1.81 & 0.13 &  29 \\
    CD Vel  &\nodata& 0.50 & 6050 & 1.70 & 3.0 & -1.70 & -1.71 & 0.12 & 102 & -1.73 & 0.08 &  22 \\
    AS Vir  & -1.49 & 0.50 & 6000 & 1.70 & 2.8 & -1.50 & -1.59 & 0.12 & 108 & -1.54 & 0.15 &  23 \\
    ST Vir  & -0.88 & 0.07 & 7650 & 2.00 & 3.3 & -0.90 & -0.80 & 0.16 &  70 & -0.79 & 0.15 &  22 \\
    ST Vir  & -0.88 & 0.29 & 6450 & 1.90 & 3.1 & -0.90 & -0.91 & 0.13 & 113 & -0.89 & 0.12 &  20 \\
    ST Vir  & -0.88 & 0.51 & 6300 & 2.20 & 2.8 & -0.90 & -0.85 & 0.15 & 100 & -0.82 & 0.19 &  22 \\
    ST Vir  & -0.88 & 0.74 & 6300 & 2.20 & 3.8 & -0.90 & -0.88 & 0.15 &  78 & -0.92 & 0.21 &  18 \\
    ST Vir  & -0.88 & 0.82 & 6350 & 2.10 & 3.7 & -0.90 & -0.81 & 0.14 &  67 & -0.89 & 0.13 &  13 \\
    UU VIr  & -0.82 & 0.25 & 6450 & 2.00 & 2.6 & -0.90 & -0.97 & 0.14 & 114 & -0.96 & 0.19 &  25 \\
    UU Vir  & -0.82 & 0.33 & 6350 & 2.00 & 2.9 & -0.90 & -0.90 & 0.10 & 112 & -0.89 & 0.12 &  22 \\
    UU Vir  & -0.82 & 0.44 & 6200 & 1.90 & 2.7 & -0.90 & -0.93 & 0.11 & 113 & -0.92 & 0.21 &  25 \\
\enddata
\end{deluxetable}
\end{center}

\begin{deluxetable}{crrrrrrrrrr}
\tabletypesize{\footnotesize}
\tablewidth{0pt}
\tablecaption{Mean Abundance Ratios\label{tab-abunds}}
\tablecolumns{11}
\tablehead{
\colhead{Star}                     &
\colhead{[Fe/H]}                   &
\colhead{[Fe/H]}                   &
\colhead{[Ca/Fe]}                  &
\colhead{[Sc/Fe]}                  &
\colhead{[Ti/Fe]}                  &
\colhead{[Ti/Fe]}                  &
\colhead{[Cr/Fe]}                  &
\colhead{[Cr/Fe]}                  &
\colhead{[Mn/Fe]}                  &
\colhead{[Ba/Fe]}                  \\
\colhead{}                         &
\colhead{{\sc i}}                  &
\colhead{{\sc ii}}                 &
\colhead{{\sc i}}                  &
\colhead{{\sc ii}}                 &
\colhead{{\sc i}}                  &
\colhead{{\sc ii}}                 &
\colhead{{\sc i}}                  &
\colhead{{\sc ii}}                 &
\colhead{{\sc i}}                  &
\colhead{{\sc ii}}                  
}
\startdata
WY Ant    & $-$1.82 & $-$1.77 &    0.32 &    0.05 &    0.39 &    0.29 &    0.07 &    0.03 & $-$0.50 & \nodata \\
BS Aps    & $-$1.51 & $-$1.44 &    0.28 & $-$0.05 &    0.27 &    0.18 &    0.00 &    0.07 & $-$0.34 & $-$0.07 \\
XZ Aps    & $-$1.65 & $-$1.65 &    0.30 &    0.05 &    0.32 &    0.35 &    0.02 &    0.10 & $-$0.31 & $-$0.07 \\
DN Aqr    & $-$1.77 & $-$1.74 &    0.36 &    0.01 &    0.34 &    0.27 &    0.03 &    0.11 & $-$0.46 & $-$0.09 \\
SW Aqr    & $-$1.39 & $-$1.36 &    0.35 & $-$0.17 &    0.29 &    0.13 &    0.02 &    0.04 & $-$0.35 & $-$0.14 \\
X Ari     & $-$2.60 & $-$2.60 &    0.35 &    0.25 &    0.45 &    0.44 & $-$0.11 &    0.21 & $-$0.75 & $-$0.68 \\
RR Cet    & $-$1.57 & $-$1.56 &    0.31 &    0.10 &    0.28 &    0.24 & $-$0.02 &    0.07 & $-$0.33 & $-$0.01 \\
W Crt     & $-$0.76 & $-$0.74 &    0.30 & $-$0.24 &    0.04 &    0.03 & $-$0.06 & $-$0.01 & $-$0.26 & $-$0.55 \\
DX Del    & $-$0.52 & $-$0.58 &    0.07 & $-$0.43 & $-$0.28 & $-$0.11 & $-$0.30 & $-$0.05 &    0.01 & $-$0.14 \\
SX For    & $-$1.80 & $-$1.79 &    0.39 & $-$0.02 &    0.29 &    0.24 & $-$0.01 &    0.07 & $-$0.46 & $-$0.13 \\
DT Hya    & $-$1.43 & $-$1.35 &    0.29 & $-$0.03 &    0.29 &    0.27 &    0.21 &    0.14 & $-$0.33 & \nodata \\
V Ind     & $-$1.62 & $-$1.59 &    0.34 & $-$0.02 &    0.27 &    0.21 &    0.05 &    0.07 & $-$0.43 & $-$0.13 \\
SSLeo     & $-$2.36 & $-$2.32 &    0.41 &    0.03 &    0.48 &    0.42 &    0.04 &    0.29 & $-$0.75 &    0.05 \\
ST Leo    & $-$1.32 & $-$1.29 &    0.32 & $-$0.10 &    0.28 &    0.23 &    0.18 &    0.16 & $-$0.44 & $-$0.03 \\
Z Mic     & $-$1.46 & $-$1.46 &    0.27 & $-$0.09 &    0.27 &    0.25 &    0.07 &    0.08 & $-$0.40 & $-$0.06 \\
RV Oct    & $-$1.47 & $-$1.44 &    0.30 & $-$0.06 &    0.29 &    0.21 &    0.09 &    0.03 & $-$0.52 & $-$0.32 \\
UV Oct    & $-$1.69 & $-$1.69 &    0.29 & $-$0.03 &    0.25 &    0.23 &    0.06 &    0.10 & $-$0.48 & $-$0.23 \\
V0445 Oph & $-$0.05 & $-$0.19 & $-$0.10 & $-$0.58 & $-$0.51 & $-$0.20 & $-$0.15 &    0.07 &    0.11 & $-$0.23 \\
AV Peg    & $-$0.14 & $-$0.20 &    0.12 & $-$0.43 & $-$0.38 & $-$0.24 & $-$0.23 & $-$0.02 & $-$0.11 & $-$0.30 \\
HH  Pup   & $-$0.95 & $-$0.90 &    0.06 & $-$0.71 & $-$0.16 & $-$0.27 & $-$0.08 &    0.02 & $-$0.08 & $-$0.33 \\
AN Ser    &    0.06 &    0.03 &    0.01 & $-$0.51 & $-$0.46 & $-$0.36 & $-$0.26 & $-$0.01 &    0.03 & $-$0.32 \\
VY Ser    & $-$1.88 & $-$1.87 &    0.35 &    0.09 &    0.30 &    0.28 & $-$0.01 &    0.08 & $-$0.51 &    0.02 \\
v1645 Sgr & $-$1.81 & $-$1.79 &    0.31 & $-$0.03 &    0.31 &    0.22 & $-$0.06 &    0.17 & $-$0.34 & $-$0.11 \\
W Tuc     & $-$1.77 & $-$1.74 &    0.35 &    0.02 &    0.31 &    0.25 & $-$0.05 &    0.15 & \nodata &    0.19 \\
CD Vel    & $-$1.71 & $-$1.73 &    0.29 & $-$0.03 &    0.31 &    0.25 &    0.19 &    0.03 & $-$0.50 & $-$0.09 \\
AS Vir    & $-$1.59 & $-$1.54 &    0.31 & $-$0.13 &    0.26 &    0.15 &    0.18 &    0.10 & $-$0.42 & $-$0.22 \\
ST Vir    & $-$0.85 & $-$0.86 &    0.20 & $-$0.24 &    0.03 &    0.06 & $-$0.08 &    0.01 & $-$0.17 & $-$0.31 \\
UU VIr    & $-$0.93 & $-$0.92 &    0.23 & $-$0.59 &    0.10 &    0.10 &    0.00 &    0.09 & $-$0.08 & $-$0.29 \\
          &         &         &         &         &         &         &         &         &         &         \\
$\langle \sigma \rangle$ &
               0.14 &    0.16 &    0.15 &    0.11 &    0.11 &    0.11 &    0.11 &    0.13 &    0.10 &    0.09 \\
$\langle \# \rangle$  &     
                 97 &      22 &       8 &       3 &       5 &       7 &       2 &       3 &       2 &       2 \\
\enddata

\end{deluxetable}

\clearpage
\begin{center}
\begin{deluxetable}{crrrrr}
\tabletypesize{\footnotesize}
\tablewidth{0pt}
\tablecaption{Radial velocity and flux measurements.\label{tab-flux}}
\tablecolumns{6}
\tablehead{
Star                  & 
[Fe/H]                & 
Vamp                  & 
DRVm\tablenotemark{a} &
DRVa\tablenotemark{a} &
H$\alpha$ flux        \\
                      &
                      &
\kmsec\               &
\kmsec\               &
\kmsec\               &
relative
}
\startdata
WY Ant    & $-$1.80  & 0.83  & 58.3  & 116.0    & 1.20\\
XZ Aps-1  & $-$1.65  & 1.10  & 63.8  & 124.5    & 1.25\\
XZ Aps-2  & \nodata  & 1.10  & 59.0  & \nodata  & 1.25\\
DN Aqr    & $-$1.76  & 0.72  & 54.0  & \nodata  & 0.44\\
SW Aqr-1  & $-$1.38  & 1.28  & 63.0  & 120.0    & 0.38\\
SW Aqr-2  & \nodata  & 1.28  & 62.2  & \nodata  & 0.38\\
X Ari-1   & $-$2.60  & 0.94  & 58.8  & 119.5    & 1.65\\
X Ari-2   & \nodata  & 0.94  & 53.7  & \nodata  & 1.65\\
RR Cet    & $-$1.57  & 0.82  & 59.4  & 115.5    & 1.12\\
W Crt     & $-$0.59  & 1.10  & 66.6  & 111.0    & 0.07\\
DX Del    & $-$0.44  & 0.70  & 53.3  & 102.0    & 0.00\\
SX For    & $-$1.80  & 0.64  & 50.0  & 101.0    & 0.18\\
V Ind     & $-$1.61  & 1.06  & 53.5  & 107.0    & 0.17\\
DT Hya    & $-$1.39  & 0.98  & 60.7  & 118.0    & 0.80\\
SS Leo-1  & $-$2.34  & 1.10  & 60.0  & 123.0    & 1.40\\
SS Leo-2  & \nodata  & 1.10  & 57.0  & \nodata  & 1.40\\
ST Leo    & $-$1.31  & 1.19  & 64.3  & 116.0    & 0.90\\
Z Mic     & $-$1.46  & 0.64  & 51.0  & 100.5    & 0.06\\
RV Oct    & $-$1.46  & 1.13  & 63.0  & 126.0    & 1.20\\
V445 Oph  & $-$0.11  & 0.81  & 56.0  &  89.0    & 0.00\\
AV Peg    & $-$0.13  & 0.96  & 62.0  &  94.0    & 0.00\\
HH Pup    & $-$0.74  & 1.35  & 69.7  & 115.5    & 0.34\\
AN Ser-1  & $+$0.05  & 1.01  & 59.4  &  97.0    & 0.32\\
AN Ser-2  & \nodata  & 1.01  & 58.1  & \nodata  & 0.32\\
VY Ser    & $-$1.88  & 0.63  & 49.0  & 101.0    & 0.20\\
W Tuc     & $-$1.74  & 1.00  & 61.0  & 131.0    & 1.20\\
CD Vel    & $-$1.72  & 0.87  & 51.0  & 104.0    & 0.39\\
ST Vir    & $-$0.86  & 1.18  & 56.2  & \nodata  & 0.00\\
ST Vir    & \nodata  & 1.18  & 64.0  & 100.5    & 0.00\\
UU Vir-1  & $-$0.93  & 1.08  & 68.0  & 118.0    & 0.96\\
UU Vir-2  & \nodata  & 1.08  & 66.0  & \nodata  & 0.96\\
\enddata

\tablenotetext{a}{typical uncertainties for DRVm are $\pm$1.5 \kmsec\
and for DRVa are $\pm$4.0 \kmsec}
\end{deluxetable}
\end{center}

\clearpage
\begin{center}
\begin{deluxetable}{cccccc}
\tabletypesize{\footnotesize}
\tablewidth{0pt}
\tablecaption{Rise time, primary \& secondary accelerations, radius-variation 
              and $\gamma$-velocity.\label{tab-rv}}
\tablecolumns{6}
\tablehead{
Star ID                             &
Rise time\tablenotemark{a}          &
Primary\tablenotemark{a}            & 
Secondary\tablenotemark{a}          & 
Radius\tablenotemark{a}             &
$\gamma$--velocity\tablenotemark{a} \\
                                    &
                                    &
acceleration                        & 
acceleration                        & 
variation                           &
                                    \\
                                    & 
percent                             & 
km s$^{-2}$                         & 
km s$^{-2}$                         & 
R$\odot$                            &
\kmsec\           
}
\startdata
\multicolumn{6}{c}{Metal-Poor RR\,ab stars} \\
X\,Ari    & 13 & 20.5 &    2.4 & 0.90 & $-$38 \\
SS\,Leo   & 12 & 22.6 &    2.3 & 0.88 & $+$162 \\
SW\,Aqr   & 11 & 32.5 &    3.8 & 0.70 & $-$50 \\
VY\,Ser   & 20 & 14.5 &\nodata & 0.80 & $-$147 \\
XZ\,Aps   & 12 & 26.1 &    2.4 & 0.84 & $+$198 \\
W\,Tuc    & 13 & 22.8 &    2.9 & 0.89 & $+$64 \\
DN\,Aqr   & 16 & 18.2 &    2.8 & 0.83 & $-$228 \\
SX\,For   & 16 & 16.2 &    2.4 & 0.70 & $+$245 \\
WY\,Ant   & 14 & 23.7 &    3.3 & 0.79 & $+$204 \\
CD\,Vel   & 14 & 20.4 &    2.9 & 0.68 & $+$241 \\
RR\,Cet   & 12 & 24.4 &    2.8 & 0.80 & $-$77\\
V\,Ind    & 15 & 19.1 &    2.8 & 0.82 & $+$201 \\
AM\,Vir   & 17 & 18.2 &    2.9 & 0.84 & $+$92 \\
RV\,Oct   & 11 & 28.3 &    3.3 & 0.79 & $+$141 \\
ST\,Leo   & 10 & 33.3 &    3.6 & 0.70 & $+$166 \\
Z\,Mic    & 18 & 17.1 &    1.3 & 0.69 & $-$60 \\
DT\,Hya   & 12 & 25.1 &    2.0 & 0.79 & $+$80 \\
\multicolumn{6}{c}{Metal-Rich RR\,ab stars} \\
UU\,Vir   & 11 & 34.6 &    4.4 & 0.73 & $-$10 \\
ST\,Vir   & 12 & 31.5 &\nodata & 0.58 & $-$4 \\
HH\,Pup   &  8 & 45.7 &    5.0 & 0.62 & $+$19 \\
DX\,Del   & 14 & 23.4 &    3.0 & 0.60 & $-$58 \\
W\,Crt    &  9 & 41.5 &    4.8 & 0.61 & $+$61 \\
V445\,Oph & 13 & 30.1 &\nodata & 0.54 & $-$21 \\
AV\,Peg  &  14 & 35.2 &    4.4 & 0.57 & $-$60 \\
AN\,Ser  &  12 & 27.5 &\nodata & 0.69 & $-$43 \\
\enddata

\tablenotetext{a}{The error bars are $\pm$1\% for rise time, 
                  $\pm$\,0.5\,km s$^{-2}$ for primary acceleration,
                  $\pm$\,1.0\,km s$^{-2}$ for secondary acceleration,
                  $\pm$\,0.02\,R$\odot$ for radius variation,
                  and $\pm$\,1.5\,km s$^{-1}$ for $\gamma$--velocity}
\end{deluxetable}
\end{center}

\end{document}